# On the Computation of Fully Proportional Representation

**Nadja Betzler**                                    NADJA.BETZLER@CAMPUS.TU-BERLIN.DE
*Institut für Softwaretechnik und*
*Theoretische Informatik*
*TU Berlin*

**Arkadii Slinko**                                    SLINKO@MATH.AUCKLAND.AC.NZ
*Department of Mathematics*
*University of Auckland*

**Johannes Uhlmann**                          JOHANNES.UHLMANN@CAMPUS.TU-BERLIN.DE
*Institut für Softwaretechnik und*
*Theoretische Informatik*
*TU Berlin*

## Abstract

We investigate two systems of fully proportional representation suggested by Chamberlin & Courant and Monroe. Both systems assign a representative to each voter so that the "sum of misrepresentations" is minimized. The winner determination problem for both systems is known to be NP-hard, hence this work aims at investigating whether there are variants of the proposed rules and/or specific electorates for which these problems can be solved efficiently. As a variation of these rules, instead of minimizing the sum of misrepresentations, we considered minimizing the maximal misrepresentation introducing effectively two new rules. In the general case these "minimax" versions of classical rules appeared to be still NP-hard.

We investigated the parameterized complexity of winner determination of the two classical and two new rules with respect to several parameters. Here we have a mixture of positive and negative results: e.g., we proved fixed-parameter tractability for the parameter the number of candidates but fixed-parameter intractability for the number of winners.

For single-peaked electorates our results are overwhelmingly positive: we provide polynomial-time algorithms for most of the considered problems. The only rule that remains NP-hard for single-peaked electorates is the classical Monroe rule.

## 1. Introduction

There is an important conceptual difference in the purpose of single-winner and multi-winner elections. Single-winner social choice rules are used to make final decisions, e.g., to elect a president or to choose a certain course of action. The multi-winner election rules are used to elect an assembly whose members will be authorized to take final decisions on behalf of the society. As a result the main property that multi-winner rules have to satisfy is that the elected assembly represents the society adequately. This, in particular, means that when a final decision is taken all opinions existing in the society are heard and taken into consideration. As Black powerfully expressed it:

> A scheme of proportional representation attempts to secure an assembly whose membership will, so far as possible, be proportionate to the volume of the different shades of political opinion held throughout the country; the microcosm is to be a true reflexion of the macrocosm (Black, 1958, p. 75).

And although any single-winner social choice rule can be easily extended to select an assembly—e.g., by choosing candidates with best scores or applying the rule repeatedly until the required quantity of representatives is elected—this is a wrong approach to the problem (Brams & Fishburn, 2002) (see also Lu & Boutilier, 2011 for some experimental evidence). The reason is that the majoritarian logic which dominates the design of single-winner social choice rules cannot provide for a balanced assembly membership.

The standard solution to the problem of electing an assembly has been the division of the election into single-member districts with approximately equal population. Each district elects one member of the assembly using a single-winner rule, normally the plurality. And although one might question whether districting should be instead based on the total adult population or on the number of registered voters, the current practice is well established and entrenched by law in many countries, including the United States (Brams, 2008). However the main problem with this approach is not the districting but the fact that it also fails to give a representation to minorities; a minority may comprise 49% of the population and be not represented in the assembly. On the positive side the districting method provides for a high level of accountability: voters who know who is their representative, can address them on particular issues and can even recall them, if they fail to represent them to a decent standard.

Various voting systems—e.g., Cumulative Voting, Single Non-Transferable vote, multi-winner variants of Single Transferable Vote, various party list systems—have been designed to solve the problem of representation of minorities (Brams & Fishburn, 2002). However none of them scored high on the accountability. It may even seem that we have a certain trade-off here and we cannot have both representation of minorities and accountability. However this is not the case. An important idea was suggested by Charles Dodgson (1884), known also as Lewis Carroll, and considered in a different form by Black (1958). Then the idea was further developed by Chamberlin and Courant (1983) and later by Monroe (1995). The relative advantages of both methods from political science point of view have been extensively discussed by Brams (2008). Dodgson asserted that a representation system should find the coalitions in the election that would have formed if the voters had the necessary time and information and allow each of the coalitions to elect their representative. If this is adopted, then a sizable minority can form a coalition and be represented.

The realization of this idea required a new concept which is the concept of *misrepresentation*. It is assumed that voters form individual preferences over the candidates based on their political ideology and "their judgement about the abilities of candidates to participate in deliberations and decision making consistent with how the individuals would wish to act were they present" (Chamberlin & Courant, 1983, p. 722). This was in a way a revolution.

Indeed, in the "single-winner" literature on voting rules it is widely accepted that voter's political preferences are more complex than their first choices alone. However, in "multi-winner" voting literature fixation on first preferences led researchers to think about proportional representation exclusively in terms of first preferences. In list systems of proportional representation, parties are assigned a number of seats in parliament that is proportional to the number of votes (first preferences) they received. The systems like Single Non-Transferable Vote, Block Voting and Cumulative Voting also do not take second preferences in account (Levin & Nalebuff, 1995). Only the Single Transferable Vote is a system of proportional representation—in fact a family of voting methods according to Tideman and Richardson (2000)—that allows voters to express the order of preference of candidates (Levin & Nalebuff, 1995). Voters rank the candidates in order of preference; first preference votes are the first to be looked at, and the votes are then transferred, if necessary, from candidates who have either been comfortably elected or who have done so badly that they have been eliminated from the election.[1]

So, if a voter is represented by a candidate who is her first preference it is reasonable to say that she is represented optimally or that the misrepresentation in this case is zero. In general, if a voter is represented by a candidate who is her $i$th preference we may assume that she is misrepresented to the degree $s_i$. Of course, it is reasonable to assume that $0 = s_1 \leq s_2 \leq \ldots \leq s_m$. So in this case the rule for measuring the total misrepresentation is fully defined by the vector $\mathbf{s} = (s_1, \ldots, s_m)$, where $m$ is the number of candidates. Using the analogy with positional scoring rules for single-winner elections we may say that this misrepresentation function is *positional*. In general, the problem of choosing a proper misrepresentation function is far from being trivial. In the work of Levin and Nalebuff (1995, p. 4) this difficulty is vividly illustrated: "if the electorate is uniformly distributed on the segment between 0 and 1, and we are to choose three representatives, should they be equally spaced $[0.25, 0.5, 0.75]$, or should they be selected so as to minimize the average distance traveled to the nearest legislator $[0.16, 0.5, 0.83]$?" In the broadest possible framework the misrepresentation function may be even voter-dependent.

Staying with the classical positional misrepresentation functions for the time being suppose that every voter is assigned to a representative in some way. Measuring the total misrepresentation for the whole society we may adopt either the Harsanyi (1995) approach or the Rawlsian one (assuming that the utility of being represented by the $i$th best candidate is $-s_i$, i.e., a nonpositive value). By Harsanyi we will have to measure the total misrepresentation as

$$M_H = \sum_{i=1}^{m} n_i s_i,$$

where $n_i$ is the number of voters represented by their $i$th most preferred candidate. According to Rawls (1999) "welfare is maximized when the utility of those society members that have the least is the greatest." This leads to the total misrepresentation function

$$M_R = \max_{i \text{ with } n_i > 0} s_i.$$

Both Chamberlin and Courant (1983), and Monroe (1995) consider that the best set of representatives must minimize the total misrepresentation which they both calculate using the Borda vector of scores, that is, $(0, 1, 2, \ldots, m - 1)$ and Harsanyi's misrepresentation function. Their methods are however different and the difference is very important. Chamberlin and Courant did not impose any restriction on the function that assigns candidates to voters. This may potentially lead to a different number of voters represented by each candidate. To remedy this Chamberlin and Courant suggested to use weighted voting in the assembly where each elected candidate has the weight equal to the number of voters they represent. Monroe rejected this approach and insisted on the principle 'one member of assembly one vote'. For this reason he insisted that the difference between the numbers of voters assigned to any two representatives is at most one.

---

1. In Northern Ireland this is the voting system used for elections to local councils, the Assembly, and the European Parliament. It is used for all elections in the Irish Republic, Malta, and Australia (although single-member constituencies are prevalent in Australia, apart from state level elections in Tasmania and the ACT). Several other countries have recently debated adopting it.

The reasons why Monroe rejected the Chamberlin and Courant approach are quite substantial. Proportional allocation of weights is known to result in excessive voting powers for the electorates of larger constituencies and there can be much debate on what is the right way of allocating weights to representatives. An alternative to Chamberlin and Courant's method would be to give to representatives weights proportional to the square root of the number of voters they represent. This is justified by the fact, that due to the square root law of Penrose (1946), the a priori voting power (as defined by the Penrose-Banzhaf index) of a member of a voting body is inversely proportional to the square root of its size. On the basis of this theory, for example, Poland insisted that the EU allocate Council-of-Minister votes according to the square root of each nation's population (Slomczynski & Zyczkowski, 2006).

The computational problems which the Harsanyi approach entails are known to be NP-hard (Lu & Boutilier, 2011; Procaccia, Rosenschein, & Zohar, 2008) for several classical misrepresentation functions. In this paper we try to achieve tractability in multiwinner elections in three different ways. Firstly, we ask whether or not the problem of finding an optimal fully proportional representation becomes easier for these classical misrepresentation functions if we adopt the Rawlsian approach for measurement of total misrepresentation. The second goal is to find the parameterized complexity of the aforementioned problems for some natural choices of parameters. The third is to develop efficient algorithms for achieving an optimal fully proportional representation in single-peaked elections.

In the remainder of this section, we formally introduce the computational problems considered in this paper, summarize the results in the extant literature, and describe our approaches and results.

## 1.1 Computational Problems Considered

An *election* is a pair $E = (C, V)$ where $C$ is a set of *candidates* (or *alternatives*) and $V$ is an ordered list of *voters*. Each voter is represented by her *vote*, which is a strict, linear order over the set of candidates (also called this voter's *preference order*). We will refer to the list $V$ as a *preference profile*, and we denote the number of voters in $V$ by $n$. The number of alternatives will be denoted by $m$. If the order of voters is not important (the election is anonymous), then $V$ can be considered as a multiset[2] of votes. In this paper we will consider only anonymous elections.

By $\mathrm{pos}_v(c)$ we will denote the position of the alternative $c$ in the ranking of voter $v$; the top-ranked alternative has position 1, the second best has position 2, etc.

---

2. This is not a set since two different voters may have the same preference order.

**Definition 1.** Given a profile $V$ over $C$, a mapping $r \colon V \times C \to \mathbb{Q}_0^+$ will be called *a misrepresentation function* if for any $v \in V$ and any two candidates $c, c' \in C$ the condition $\mathrm{pos}_v(c) < \mathrm{pos}_v(c')$ implies $r(v, c) \le r(v, c')$.

This is to say that if $c$ is preferred to $c'$ in $v$'s ranking, then the misrepresentation of $v$, when she is represented by $c'$ will be at least as large as her misrepresentation, when she is represented by $c$. In the classical framework the misrepresentation of a candidate for a voter is a function of the position of the candidate in the preference order of that voter given by $\mathbf{s} = (s_1, \ldots, s_m)$, i.e., the misrepresentation function in this case will be

$$r(v, c) = s_{\mathrm{pos}_v(c)}.$$

An important particular case is the *Borda misrepresentation function* defined by the vector $(0, 1, \ldots, m - 1)$.

In the approval voting framework, if a voter is represented by a candidate whom she approves, her misrepresentation is considered to be zero, otherwise it is equal to one. This function is called the *approval misrepresentation function*. This misrepresentation function does not have to be positional since different voters may approve different number of candidates. Note that some misrepresentation functions, like Borda, can be derived from the preference lists of the voters. In contrast, an approval misrepresentation function cannot be obtained from a preference list without further information about the threshold that separates approved candidates from disapproved ones. In the general framework the misrepresentation function may be arbitrary.

By $w \colon V \to C$ we denote the function that assigns voters to representatives (or the other way around), i.e., under this assignment voter $v$ is represented by candidate $w(v)$. The total misrepresentation of the election under $w$ is then given by

$$\sum_{v \in V} r(v, w(v)) \quad \text{or} \quad \max_{v \in V} r(v, w(v))$$

in the Harsanyi's classical and Rawls' minimax versions, respectively. We say that a mapping $w$ respects the *M-criterion* (or *Monroe criterion*) if $|w(V)| = k$ and $w$ assigns at least $\lfloor n/k \rfloor$ and at most $\lceil n/k \rceil$ voters to every candidate from $w(V)$, where $k$ is the total number of representatives to be elected to the assembly. Note that in case of the *M*-criterion a set of more than $k$ winners might lead to a higher misrepresentation than a set of $k$ winners. For example, consider an election such that all voters favour the same candidate but the set of winners that has to be elected is greater than one.

Based on the previous discussion, in this work we investigate the computational complexity of the following four combinatorial problems. The two classical ones described above are named after Chamberlin and Courant (CC), for the case when a candidate can represent an arbitrary number of voters (and this number of voters will be his weight in the elected assembly), and Monroe (M), for the case when every candidate represents roughly the same number of voters (and each representative has one vote in the assembly). The two previously unstudied versions which adopt the Rawlsian approach for measuring the total misrepresentation are called the *minimax* versions of the classical ones.

### CC-MULTIWINNER (CC-MW)

**Given:** A set $C$ of candidates, a multiset $V$ of voters, a misrepresentation function $r$, a misrepresentation bound $R \in \mathbb{Q}_0^+$ and a positive integer $k$.

**Task:** Find a subset $C' \subseteq C$ of size $k$ and an assignment of voters $w$ such that $w(V) = C'$ and $\sum_{v \in V} r(v, w(v)) \leq R$.

MINIMAX CC-MULTIWINNER (MINIMAX CC-MW)

**Given:** A set $C$ of candidates, a multiset $V$ of voters, a misrepresentation function $r$, a misrepresentation bound $R \in \mathbb{Q}_0^+$ and a positive integer $k$.

**Task:** Find a subset $C' \subseteq C$ of size $k$ and an assignment of voters $w$ such that $w(V) = C'$ and $\max_{v \in V} r(v, w(v)) \leq R$.

M-MULTIWINNER (M-MW)

**Given:** A set $C$ of candidates, a multiset $V$ of voters, a misrepresentation function $r$, a misrepresentation bound $R \in \mathbb{Q}_0^+$ and a positive integer $k$.

**Task:** Find a subset $C' \subseteq C$ of size $k$ and an assignment of voters $w$, which respects the $M$-criterion, $w(V) = C'$ and such that $\sum_{v \in V} r(v, w(v)) \leq R$.

MINIMAX M-MULTIWINNER (MINIMAX M-MW)

**Given:** A set $C$ of candidates, a multiset $V$ of voters, a misrepresentation function $r$, a misrepresentation bound $R \in \mathbb{Q}_0^+$ and a positive integer $k$.

**Task:** Find a subset $C' \subseteq C$ of size $k$ and an assignment of voters $w$, which respects the $M$-criterion, $w(V) = C'$ and such that $\max_{v \in V} r(v, w(v)) \leq R$.

Note that finding an assignment of voters to a fixed set of $k$ winners can be accomplished in polynomial time for all four problems by applying network flow algorithms (see Section 3.2). Hence, in what follows we assume that $k < m$ and $k < n$ since otherwise all four problems can be decided in polynomial time. We also note that all problems considered are contained in NP since one can guess a set of $k$ winners and a corresponding "mapping" to the voters and check in polynomial time whether it satisfies the corresponding conditions.

The four problems above are stated for general misrepresentation functions (since some of our algorithmic results hold even for this case) but the main focus of this work is on the Borda and the approval ones.

## 1.2 Previous Computational Complexity Results

The study of the computational complexity of problems in the context of voting was initiated by Bartholdi III, Tovey, and Trick (1989) about 20 years ago but has became an active area of research only recently (Conitzer, 2010; Faliszewski & Procaccia, 2010; Faliszewski, Hemaspaandra, & Hemaspaandra, 2010; Faliszewski, Hemaspaandra, Hemaspaandra, & Rothe, 2009b). While there is a large number of papers dealing with single-winner elections or multi-winner elections whose final goal is still to choose a single winner after a tiebreaking, only few articles (Potthof & Brams, 1998; Procaccia et al., 2008; Lu & Boutilier, 2011) deal with the computational complexity of Multiwinner elections aimed at achieving a proportional representation. In particular, these works contain NP-hardness proofs for CC-MULTIWINNER and M-MULTIWINNER for approval misrepresentation function (Procaccia et al., 2008) and for CC-MULTIWINNER for Borda misrepresentation function (Lu & Boutilier, 2011). Algorithmic approaches comprise Integer Linear Programming (Potthof &

Brams, 1998; Brams, 2008) for CC-MW and M-MW, approximation algorithms based on greedy strategies (Lu & Boutilier, 2011) for CC-MW, and polynomial-time algorithms for CC-MW and M-MW for instances where the number of candidates is constant (Procaccia et al., 2008). In contrast, to the best of our knowledge, the computational complexity of the minimax versions of the problems remained unstudied.

We are only aware of one further work explicitly studying computational complexity issues in the context of multiwinner elections. Meir, Procaccia, Rosenschein, and Zohar (2008) investigate the computational complexity of strategic voting for several multiwinner elections for which a winner can be determined in polynomial time. The systems considered do not lead to any kind of proportional representation.

## 1.3 Our Approach and Results for General Elections

Our first result is that the minimax versions of the classical Chamberlin-Courant and Monroe problems are also NP-complete. In other words, adopting the Rawlsian approach does not make computation of the problems easier in general (but we will see that the situation changes completely for single-peaked elections where the minimax version becomes easier indeed). Based on these negative results, this work aims at extending the previous algorithmic approaches described above by an analysis whether or not there are settings in which the problems become tractable. To this end, parameterized algorithmics is an appropriate tool as it aims at identifying tractable special cases of NP-hard problems. The cornerstone of this approach is the idea that the complexity of a problem is not only measured in the total size of an input instance $I$ but also in an additional parameter $p$, usually a nonnegative integer (but it can be a pair of integers or virtually anything). A problem is called *fixed-parameter tractable* if there is an algorithm solving every instance of it in $f(p) \cdot \text{poly}(|I|)$ time, where $f$ is a computable function (Downey & Fellows, 1999; Flum & Grohe, 2006; Niedermeier, 2006). For small values of $p$ an algorithm with such running time might represent an efficient algorithm for the NP-hard problem under consideration. Parameterized complexity also provides a tool of "parameterized reductions" by which one can show that a problem is presumably not fixed-parameter tractable. One of the most important parameterized complexity classes for this purpose is $W[2]$ (see Section 2 for more details). We remark in passing that a parameterized complexity analysis has been employed for several other voting problems, (e.g., see Brandt, Brill, & Seedig, 2011; Betzler, Guo, & Niedermeier, 2010; Betzler, Hemmann, & Niedermeier, 2009; Christian, Fellows, Rosamond, & Slinko, 2007; Dorn & Schlotter, 2010; Elkind, Faliszewski, & Slinko, 2010b; Faliszewski, Hemaspaandra, Hemaspaandra, & Rothe, 2009a and also Betzler, Bredereck, Chen, & Niedermeier, 2012 for a survey).

In the context of multiwinner elections, a parameter that immediately attracts attention is the number $k$ of winners, which in many settings might be much smaller than the number of candidates or the number of voters. Another reasonable parameter is the misrepresentation bound $R$ since in an ideal (or fully personalizable Lu & Boutilier, 2011) situation $R$ is equal to zero, that is, every voter is represented by one of her most preferred candidates. We provide a parameterized complexity analysis of all four considered problems for the Borda and approval misrepresentation functions with respect to the parameters $k$ and $R$.

| Parameter | $r$ | CC-MW | MINIMAX CC-MW | M-MW | MINIMAX M-MW |
|---|---|---|---|---|---|
| #winner $k$ | A | W[2]-hard ($\diamondsuit$) | W[2]-hard ($\diamondsuit$) | W[2]-hard ($\diamondsuit$) | W[2]-hard ($\diamondsuit$) |
| #winner $k$ | B | W[2]-hard ($\triangle$) | W[2]-hard ($\triangle$) | W[2]-hard ($\triangle$) | W[2]-hard ($\triangle$) |
| misr. $R$ | A | NP-h for $R = 0$ ($\diamondsuit$) | NP-h for $R = 0$ ($\diamondsuit$) | NP-h for $R = 0$ ($\diamondsuit$) | NP-h for $R = 0$ ($\diamondsuit$) |
| misr. $R$ | B | XP ($\clubsuit$) | NP-h for $R \geq 1$ ($\natural$) | XP ($\clubsuit$) | NP-h for $R \geq 1$ ($\natural$) |
| | | | P for $R = 0$ ($\sharp$) | | P for $R = 0$ ($\sharp$) |
| $(R, k)$ | A | W[2]-hard ($\diamondsuit$) | W[2]-hard ($\diamondsuit$) | W[2]-hard ($\diamondsuit$) | W[2]-hard ($\diamondsuit$) |
| $(R, k)$ | B | FPT ($\heartsuit$) | FPT ($\heartsuit$) | FPT ($\spadesuit$) | FPT ($\dagger$) |
| # cand. | U | FPT ($\natural$) | FPT ($\natural$) | FPT ($\natural$) | FPT ($\natural$) |
| # voters | U | FPT ($\flat$) | FPT ($\flat$) | FPT ($\flat$) | FPT ($\flat$) |

Table 1: Parameterized complexity of the considered multiwinner problems for instances where the misrepresentation function $r$ is either approval (A), Borda (B) or unrestricted (U). Results are obtained as follows. $\diamondsuit$: Theorem 1, $\triangle$: Theorem 2, $\clubsuit$: Theorem 3, $\natural$: Theorem 4, $\heartsuit$: Theorem 5, $\spadesuit$: Theorem 6, $\dagger$: Theorem 7 $\flat$: Proposition 1, $\sharp$ Proposition 2.

In addition, we also investigate the composite parameter $(R, k)$ consisting of the number of winners and the misrepresentation bound.

An overview of the results is provided in Table 1. When the number of winners $k$ is a parameter, all considered problems turn out to be W[2]-hard. For the parameterization by the total misrepresentation bound $R$ the results are more varied. For the case $R = 0$, for the approval misrepresentation function all four problems are NP-hard while they are solvable in polynomial time for the Borda misrepresentation function. However, MINIMAX CC-MW and MINIMAX M-MW become NP-hard for every $R \geq 1$. In contrast, the sum-minimization variants CC-MW and M-MW for the Borda misrepresentation function are solvable in polynomial time for constant $R$ (the corresponding parameterized complexity class is called XP). Note that the provided algorithm shows the containment in XP with respect to $R$ but not fixed-parameter tractability, this problem remains open. This inspired our analysis of the composite parameter $(R, k)$, covering scenarios in which there is a small set of winners that can represent all voters with a small total misrepresentation. While for the approval misrepresentation function, this still leads to parameterized intractability, for the Borda misrepresentation function, we show fixed-parameter tractability for all considered problems. To complete the picture of a multivariate complexity analysis, we additionally provide fixed-parameter tractability with respect to the parameters "number of voters" and "number of candidates".

## 1.4 Results for Single-Peaked Elections

Single-peakedness is one of the central notions in social choice and political science alike (Black, 1958; Moulin, 1991; Tideman, 2006). The preferences of voters are single-peaked when a single issue dominates their formation. This could be their ideological position on the Left-Right or Liberal-Conservative spectra, level of taxation, immigration quota, etc. Tideman compares single-peakedness with convexity of preferences and discusses when it is reasonable to assume this. He refers to a data collection containing 87 ranked-ballot real-life elections, which he has access to, and claims that most of them are single-peaked.

| CC-MW | MINIMAX CC-MW | M-MW | MINIMAX M-MW |
|---|---|---|---|
| $O(nm^2)$ ◇ | $O(nm)$ ♣ | $O(n^3m^3k^3)$ for approval ♠ | $O(n^3m^3k^3)$ ♡ |
| | | NP-h for integer mis. func. △ | |

Table 2: Overview of the computational complexity for singled-peaked elections. In the case of polynomial-time solvability, the table provides the running times depending on the number $n$ of voters, the number $m$ of candidates, and the number $k$ of winners. If not stated otherwise, the result holds for an arbitrary misrepresentation function. ◇: Theorem 8, ♣: Proposition 4, ♠: Theorem 10, ♡: Proposition 5, △: Theorem 11.

This single dominating issue is normally represented by an axis and each voter is characterized by a single point on this axis (see an example on Figure 1. The misrepresentation function for a fixed voter is then a function of a single variable defined on that axis. The single-peakedness of preferences implies that this function has exactly one local minimum. We refer to Section 5 for a formal proof of this statement.

We note that for votes in the form of approval ballots as well as linear orders, single-peakedness of the profile can be checked in linear time (Booth & Lueker, 1976; Escoffier, Lang, & Öztürk, 2008) with the reconstruction of the order of the candidates on the axis.

In the case of single-peaked profiles some computational problems have turned out to allow for more efficient solving strategies than in the general case (Brandt, Brill, Hemaspaandra, & Hemaspaandra, 2010; Conitzer, 2009). In particular, the study of the computational complexity of voting rules with NP-hard winner-determination problem shows that for all Condorcet-consistent ones—and these include Dodgson, Kemeny, and Young rules—the winner-determination problem becomes polynomial-time solvable if we restrict ourselves to single-peaked profiles (Brandt et al., 2010). The obvious reason for this is that single-peakedness eliminates the possibility of Condorcet cycles in the election profile.

It is not that obvious that single-peakedness must also simplify the winner-determination problem for methods of proportional representation. However, it seems natural to investigate this possibility. Our results show that in many instances the winner-determination problem for methods of proportional representation does indeed become easier too.

Our results are summarized in Table 2. For CC-MW and MINIMAX CC-MW the problems are solvable in polynomial time for an arbitrary misrepresentation function. More specifically, for CC-MW we provide a dynamic programming algorithm running in $O(nm^2)$ time for $n$ voters and $m$ candidates, and MINIMAX CC-MW can be solved in $O(nm)$ time by a greedy algorithm. For the Monroe system and its variants, the results become more diverse. While MINIMAX M-MW for the general misrepresentation function is still solvable in polynomial time, M-MW is NP-hard. However, on the positive side, we can still show polynomial-time solvability for M-MW for the approval misrepresentation function. Basically, the results are obtained as follows. For M-MW for the approval misrepresentation function we establish a close connection to a "one-dimensional rectangle stabbing" problem with capacities. This allows to provide a dynamic programming algorithm based on a decomposition property provided by Even, Levi, Rawitz, Schieber, Shahar, and Sviridenko (2008). This result can be transferred to MINIMAX M-MW. The NP-hardness of M-MW is established by a many-one reduction from a restricted version of the EXACT 3-COVER problem. The NP-hardness holds for an integer-valued misrepresentation function for which the maximum misrepresentation value is still polynomial in the number of candidates. However, we need to allow situations in which a voter may be equally misrepresented by several candidates. Hence, it is not clear how to transfer the corresponding many-one reduction to M-MW for the Borda misrepresentation function. For this problem the computational complexity is left open.

## 1.5 Organization of the Paper

The paper is organized as follows. In Section 2, we introduce the main concepts of parameterized complexity and some graph problems. Section 3 contains basic observations about the relations of the four problems under consideration and fixed-parameter tractability results with respect to the number of voters and the number of candidates. The two main contributions are proved in Section 4 and Section 5. In Section 4, we present our main parameterized complexity results as well as the NP-hardness results for the minimax versions. In Section 5, the special case of single-peaked elections is handled. Finally, in Section 6 we conclude with a discussion of the relevance of our results and some related problems and settings.

## 2. Preliminaries

We briefly introduce the framework of parameterized complexity followed by some basic graph problems that are employed in this paper. For basic notions regarding classical complexity theory we refer to Garey and Johnson (1979).

### 2.1 Parameterized Complexity

The concept of parameterized complexity was pioneered by Downey and Fellows (1999). See also the textbooks by Flum and Grohe (2006) and Niedermeier (2006). The fundamental goal is to find out whether the seemingly unavoidable combinatorial explosion, occurring in exact algorithms for NP-hard problems, can be confined to certain problem-specific parameters. The idea is that when such a parameter in a real-life application is restricted to small values only, then an algorithm with a running time that is exponential exclusively with respect to the parameter may be efficient and practical. We now provide formal definitions.

**Definition 2.** A parameterized problem is a language $L \subseteq \Sigma^* \times \Sigma^*$, where $\Sigma$ is a finite alphabet. The second component is called the parameter of the problem.

Basically, this means that an input to a parameterized problem is a pair $(x, p)$, where $x$ can be considered as the "main input" and $p$ is a parameter of the problem. We consider parameters which are positive integers or "composite" parameters that are tuples of several positive integers.

**Definition 3.** A parameterized problem $L$ is fixed-parameter tractable if there is an algorithm that decides in $f(p) \cdot |x|^{O(1)}$ time whether $(x, p) \in L$, where $f$ is an arbitrary computable function that depends only on $p$. The complexity class of all fixed-parameter tractable problems is called FPT.

Unfortunately, not all parameterized problems are fixed-parameter tractable. To this end, Downey and Fellows (1999) developed a theory of parameterized intractability by means of a completeness program with complexity classes. More specifically, they defined the so-called $W$-hierarchy by using Boolean circuits. This hierarchy consists of the following classes:

$$\text{FPT} \subseteq \text{W}[1] \subseteq \text{W}[2] \subseteq \ldots \subseteq \text{W}[\text{Sat}] \subseteq \text{W}[\text{P}] \subseteq \text{XP}$$

(we refer the reader to the book by Downey & Fellows, 1999, for their precise description).

In particular, we stress that the concept of fixed-parameter tractability is different from the notion of "polynomial-time solvability for constant $p$" since an algorithm running in $O(|x|^p)$ time does not imply fixed-parameter tractability. All problems that can be solved in the running time $O(|x|^{f(p)})$ for a computable function $f$ form the complexity class called XP.

The containment $\text{W}[1] \subseteq \text{FPT}$ would not imply P = NP as such. It would imply, however, the failure of the Exponential Time Hypothesis (Impagliazzo, Paturi, & Zane, 2001). Hence, it is commonly believed that W[1]-hard problems are not fixed-parameter tractable. To show the W[$t$]-hardness of a problem for some positive integer $t$, the following reduction concept was introduced.

**Definition 4.** Let $L, L' \subseteq \Sigma^* \times \Sigma^*$ be two parameterized problems. We say that $L$ reduces to $L'$ by a parameterized reduction if there are two computable functions $h_1 \colon \Sigma^* \to \Sigma^*$ and $h_2 \colon \mathbb{N} \to \mathbb{Q}^+$ and a function $f \colon \Sigma^* \times \Sigma^* \to \Sigma^* \times \Sigma^*$ such that for each $(x, p) \in \Sigma^* \times \Sigma^*$

1. $(x, p) \in L \iff f(x, p) \in L'$ and $f$ is computable in time $|x|^{O(1)} \cdot h_2(|p|)$ and

2. if $(x', p') = f(x, p)$, then $p' = h_1(p)$.

Analogously to the case of NP-hardness, for any positive integer $t$, it suffices to give a parameterized reduction from one W[$t$]-hard parameterized problem $X$ to a parameterized problem $Y$ to show the W[$t$]-hardness of $Y$. For more details about parameterized complexity theory we refer to the textbooks (Downey & Fellows, 1999; Flum & Grohe, 2006; Niedermeier, 2006).

In this work, we only provide results regarding the second level of (presumable) parameterized intractability captured by the complexity class W[2]. Several parameterized reductions in this work are from the W[2]-complete HITTING SET (HS) problem: Given a family $\mathcal{F} = \{F_1, \ldots, F_n\}$ of sets over a universe $U = \{u_1, \ldots, u_m\}$ and an integer $k \geq 0$, decide whether there is a *hitting set* $U' \subseteq U$ of size at most $k$ by which we understand a set $U'$ such that $F_i \cap U' \neq \emptyset$ for every $1 \leq i \leq n$. HS is NP-hard (Garey & Johnson, 1979) and W[2]-hard with respect to parameter $k$ (Downey & Fellows, 1999).

## 2.2 Graph Problems

Some of our algorithmic results employ algorithms for basic graph problems defined in the following. An *undirected graph* is a pair $G = (U, E)$, consisting of the set $U$ of vertices and the set $E$ of edges, where an edge is an unordered pair (size-two set) of vertices. Two vertices $u, v \in U$ are called *adjacent* if $\{u, v\} \in E$. For an undirected graph $G = (U, E)$ and a vertex $u \in U$, the *neighborhood* $N(u)$ of $u$ is the set of all vertices adjacent to $u$.

An undirected graph $G = (U, E)$ is called *bipartite* if the vertex set $U$ can be partitioned into two nonintersecting subsets $U_1$ and $U_2$ such that $E \subseteq \{\{u, v\} \mid u \in U_1 \text{ and } v \in U_2\}$. A *matching* is an edge set $E' \subseteq E$ such that $e \cap e' = \emptyset$ for every two distinct edges $e, e' \in E'$. A *maximum matching* is a matching with maximum cardinality. In an undirected graph where each edge $\{u, v\}$ is associated with a weight $w(\{u, v\})$ a *maximum-weight matching* is a matching $E'$ such that $\sum_{\{u,v\} \in E'} w(\{u, v\})$ is maximal.

A directed graph or a directed network is a pair $G = (U, A)$, consisting of the set $U$ of vertices and the set $A \subseteq U \times U$ of directed edges (or arcs) where each directed edge is an ordered pair of vertices. A *flow network* is a directed network $G = (U, A)$ with two distinguished vertices $s \in U$ (the source) and $t \in U$ (the sink or target) where each arc $(u, v) \in A$ is associated with a nonnegative number $c(u, v)$, called capacity. Roughly speaking, a flow is a function $f$ that assigns a real value $f(u, v)$ with $0 \leq f(u, v) \leq c(u, v)$ to every arc $(u, v) \in A$ and satisfies the constraints that for every vertex $v$ except for the source and the sink the total flow into $v$ equals the total flow out of $v$. See the textbook of Cormen, Leiserson, Rivest, and Stein (2001) for details. A *maximum flow* is a flow such that the total flow into the sink is maximal.

In this paper, we make use of the fact that a maximum-weight matching in a bipartite graph as well as a maximum flow in general graphs can be computed in polynomial time by standard graph algorithms (e.g., see Cormen et al., 2001).

## 3. Basic Results and Observations

In this section, we shed light on the combinatorial relations between the problems and investigate the parameterized complexity of the considered problems with respect to the parameters "number of voters" and "number of candidates". These results will also be employed in the following sections. In particular, some of the algorithms showing fixed-parameter tractability will be used as "subroutines" in Section 4 to obtain fixed-parameter tractability with respect to other parameterizations.

### 3.1 Relations between the Problems

Although all four problems come with different properties in general, in some special cases, some of them coincide. One such example is the so-called *fully personalizable setting* (Lu & Boutilier, 2011), that is, the case when the misrepresentation bound $R$ is zero and hence every voter has to be represented by one of her best alternatives (i.e., one of those for which the misrepresentation is zero). Clearly, asking for a set of winners and an assignment for which the sum of misrepresentations is zero is equivalent to asking for a set of winners and an assignment for which the maximum misrepresentation value is zero. This leads to the following observation.

**Observation 1.** *For $R = 0$,* MINIMAX M-MULTIWINNER *coincides with* M-MULTIWINNER *and* MINIMAX CC-MULTIWINNER *coincides with* CC-MULTIWINNER.

Moreover, for the two minimax versions of the problems, it only matters whether a particular misrepresentation value exceeds the threshold $R$ or not. Hence, an instance of a minimax version with an arbitrary misrepresentation function $r$ can be reduced to an equivalent instance of the same problem with the approval misrepresentation function $r'$ as follows. For every voter $v$ and every candidate $c$, set $r'(v, c) = 1$ if $r(v, c) > R$, and $r'(v, c) = 0$ if $r(v, c) \leq R$ and, finally, set $R' := 0$.

**Observation 2.** *For a* Minimax M-/CC-Multiwinner *instance* $(C, V, r, R, k)$ *with misrepresentation function $r$, there is an instance* $(C, V, r', 0, k)$ *with the* approval *misrepresentation function $r'$ such that the new instance is a yes-instance if and only if the original instance is a yes-instance.*

As a direct consequence, for the minimax versions every algorithm for the approval misrepresentation function also applies to instances with general misrepresentation function. Moreover, hardness results for an arbitrary misrepresentation function transfer to the approval misrepresentation function. Combining Observations 1 and 2, we conclude that an algorithm for M-MW (CC-MW) for instances with $R = 0$ also solves the corresponding minimax version for general misrepresentation function.

Finally, observe that a hardness result established for the approval misrepresentation function can be directly transferred to the minimax version of the same problem if the misrepresentation function is such that a voter is allowed to give an arbitrary number of candidates a misrepresentation value of at most $R$. Note that this does not hold for the Borda misrepresentation function where every voter $v$ must specify exactly $R+1$ candidates that can represent $v$ with misrepresentation at most $R$.

### 3.2 The Numbers of Voters and Candidates as Parameters

We argue that all four problems considered are fixed-parameter tractable with respect to the number of candidates as well as with respect to the number of voters. Our algorithms are based on brute-force search combined with maximum flow and matching techniques. First, we consider the parameterization by the number of voters. Then, we focus on the parameterization by the number of candidates.

#### 3.2.1 The Number of Voters as Parameter

We show that all considered multiwinner problems are fixed-parameter tractable when parameterized by the number $n$ of voters. The basic idea is that an assignment of candidates to voters induces a partition of the set of voters such that all voters in any set of this partition are represented by the same candidate. Given a partition of the set of voters, the best set of candidates for this partition can be found by the computation of a matching in a bipartite auxiliary graph. Since we may try all $O(k^n) \subseteq O(n^n)$ partitions of the set of voters into $k$ sets the resulting algorithm shows fixed-parameter tractability.

**Proposition 1.** (Minimax) CC-Multiwinner *and* (Minimax) M-Multiwinner *can be solved in $n^n \cdot \text{poly}(n, m)$ time for an instance with $n$ voters and $m$ candidates.*

*Proof.* First, we present a solution strategy for Minimax CC-MW. To find a set of $k$ winners, try all $O(k^n)$ partitions of the set of voters into $k$ subsets. In the case of a yes-instance of Minimax CC-MW, there must be a partition $V_1, \ldots, V_k$ of the multiset of voters $V$ as follows. For every $V_i$, all voters of $V_i$ are assigned to the same candidate $c$ of an optimal set of $k$ winners and no other voter is assigned to $c$. Hence, for every partition, it remains to select $k$ candidates, one candidate $c_i$ for every subset $V_i$, such that by assigning the voters in $V_i$ to $c_i$ the misrepresentation of any voter is at most $R$. For MINIMAX CC-MW the set of candidates can be determined by computing a maximum-cardinality matching in the following bipartite graph. One part of the graph represents the set of candidates and the other part the set $\{V_1, \ldots, V_k\}$. Moreover, there is an edge between a vertex representing a candidate $c$ and a vertex representing a subset $V_i$ if and only if $r(v, c) \leq R$ for all $v \in V_i$. It is straightforward to verify that all voters can be represented with maximal misrepresentation bound $R$ if and only if there is a maximum-cardinality matching of size $k$ (all vertices representing the subsets are "matched") in the constructed graph.

Regarding the running time, the computation of a maximum-weight matching in a bipartite graph with $n_v$ vertices and $n_e$ edges can be accomplished in $O(n_v(n_e + n_v \cdot \log n_v))$ time (Fredman & Tarjan, 1987). Since the number of edges and vertices in the constructed bipartite graph are polynomial in the number of candidates and $k \leq n$, the claimed running time follows.

Next, we focus on CC-MW. Again, we try all partitions of the voters into $k$ subsets. For every such partition, we compute a maximum-weight matching in the following edge-weighted bipartite graph. One part consists of vertices corresponding to candidates and the other part of vertices corresponding to the subsets of the partition $V_1, \ldots, V_k$ of the multiset of voters. Moreover, there is an edge between every vertex $c$ and every vertex $V_i$ with weight $T - \sum_{v \in V_i} r(v, c)$, where $T$ is a positive integer that is large enough to ensure that all weights are positive. The crucial observation is that in a maximum-weight matching every vertex from $V_1, \ldots, V_k$ is matched since the edge weights are positive (here we assume that $k \leq m$). Hence, the computation of a maximum-weight matching yields a set of $k$ candidates "representing" the subsets of voters as good as possible. More specifically, let $W$ denote the weight of the maximum-weight matching. Then, $kT - W$ is the total misrepresentation under the corresponding assignment.

Finally, observe that for the two problems where the assignment of the voters to the winners must fulfill the $M$-criterion we can proceed in the same way with the single exception that we need only to consider partitions such that every subset contains at least $\lfloor n/k \rfloor$ and at most $\lceil n/k \rceil$ voters. The running time bound follows in complete analogy to MINIMAX CC-MW as discussed above. $\qquad\square$

### 3.2.2 THE NUMBER OF CANDIDATES AS PARAMETER

For a fixed number of candidates all four considered multiwinner problems can be solved efficiently. For (MINIMAX) CC-MULTIWINNER parameterized by the number $m$ of candidates fixed-parameter tractability is trivial: We can test all $\binom{m}{k} \leq 2^m$ subsets of candidates and report a set of candidates with minimum total misrepresentation. To this end, one assigns every voter $v$ to the candidate of the considered subset that represents $v$ in the best possible way and then directly obtains the sum of misrepresentations or maximum misrepresentation.

Clearly, such an assignment of the voters does not have to fulfill the $M$-criterion. However, for (MINIMAX) M-MULTIWINNER, one can apply network flow algorithms to find an optimal assignment of the voters to a size-$k$ subset $C'$ of the set of candidates (see the Preliminaries in Subsection 2.2 for basic definitions regarding network flows).

For MINIMAX M-MULTIWINNER, construct a directed network with a vertex for every candidate from $C'$, one vertex for every voter, a source, and a sink vertex. There is an arc with capacity $\lceil n/k \rceil$ and lower bound of $\lfloor n/k \rfloor$ from the source to every "candidate-vertex"[3]. Moreover, there is a capacity-one arc from a "candidate-vertex" to a "voter-vertex" if and only if the corresponding candidate can represent the corresponding voter with misrepresentation at most $R$. Finally, there is an arc with capacity one from every "voter-vertex" to the sink vertex. It is straightforward to verify that there is a network flow of size $n$ if and only if there is an assignment from the voters to $C'$ that satisfies the $M$-criterion and every voter is represented with misrepresentation at most $R$.

For M-MULTIWINNER, the construction given for the minimax version can be further extended. In particular, it follows from Theorem 2 by Procaccia et al. (2008), that finding an $M$-criterion fulfilling assignment from $V$ to $C'$ with minimum total misrepresentation can be accomplished in polynomial time by the computation of a transportation problem or, equivalently, by the computation of a minimum-weight maximum flow.

**Proposition 2.** (MINIMAX) CC-MULTIWINNER *and* (MINIMAX) M-MULTIWINNER *can be solved in* $O(2^m \cdot nm)$ *and* $O(2^m \cdot \mathrm{poly}(n,m))$ *time, respectively, for instances with* $m$ *candidates.*

## 4. The Number of Winners and the Misrepresentation Bound as Parameters

In this section, we show that all four problems for the approval and the Borda misrepresentation functions are W[2]-hard with respect to the number of winners. For both misrepresentation functions we provide one parameterized reduction that works for all four problems. We further investigate the misrepresentation bound $R$ as parameter. While for the approval misrepresentation function NP-hardness for $R = 0$ follows directly from the parameterized reduction with respect to the number of winners, for the Borda misrepresentation function, this parameter needs to be investigated separately. We show that CC-MW and M-MW are in XP with respect to $R$, that is, they are solvable in polynomial time when $R$ is constant. However, the corresponding algorithm does not show fixed-parameter tractability with respect to $R$. The question whether or not our result can be extended to fixed-parameter tractability with respect to $R$ is left open. We present, however, some fixed-parameter tractability results with respect to the composite parameter $(R, k)$ at the end of this section. An overview of the results can be found in Table 1.

### 4.1 The Approval Misrepresentation Function

We provide a reduction from the W[2]-complete HITTING SET problem to establish W[2]-hardness for all four problems. Before doing so, we discuss some related results. In the conference paper (Procaccia, Rosenschein, & Zohar, 2007) it was stated that the NP-hardness for CC-MULTIWINNER and M-MULTIWINNER follows from a reduction from MAX $k$-COVER (omitting the problem definition and the construction) but in the subsequent journal paper (Procaccia et al., 2008) the reduction was given from EXACT 3-COVER. Although this is sufficient to show NP-hardness, a reduction from EXACT 3-COVER does not imply W[2]-hardness. The reduction given here is conceptually similar but requires some additional voters to deal with the fact that the sets of a HITTING SET instance might come with different/unbounded size.

---

3. The problem variant with lower bounds (demands) can be solved in polynomial time by a simple reduction to the normal flow problem (Ahuja, Magnanti, & Orlin, 1993, Section 6.7).

**Theorem 1.** *For the approval misrepresentation function,* (MINIMAX) CC-MULTIWINNER *and* (MINIMAX) M-MULTIWINNER *are W[2]-hard with respect to the number $k$ of winners even if $R = 0$.* MINIMAX CC-MULTIWINNER *and* MINIMAX M-MULTIWINNER *are NP-complete.*

*Proof.* First, we show W[2]-hardness for M-MULTIWINNER. Then, we argue that the presented reduction works for the other three problems as well.

Given an instance of Hitting Set $(\mathcal{F} = \{F_1, \ldots, F_n\}, U = \{u_1, \ldots, u_m\}, k)$, build an instance of M-MULTIWINNER with set $C$ of candidates as follows. There is a candidate $c_i \in C$ for every element $u_i \in U$. The multiset of voters is $V_{\mathcal{F}} \uplus D$, where $V_{\mathcal{F}} := \{v_F \mid F \in \mathcal{F}\}$ is the multiset of voters indexed by $\mathcal{F}$ and $|D| = n(k-1)$ is a set of dummy voters. Furthermore, for every $F \in \mathcal{F}$ and every $u_i \in U$, let $r(v_F, c_i) := 0$ if $u_i \in F$ and $r(v_F, c_i) := 1$, otherwise. Finally, for every $d \in D$ and every $u_i \in U$, set $r(d, c_i) := 0$. This completes the construction. For the correctness we show the following.

> *Claim.* There is a hitting set of size $k$ for $\mathcal{F}$ if and only if there is a winner set of size $k$ for M-MULTIWINNER that represents all voters with total misrepresentation $R = 0$.

"$\Rightarrow$": Let $U'$ denote a size-$k$ hitting set for $\mathcal{F}$ and $C' := \{c_i \mid u_i \in U'\}$. We show that one can build a mapping $w : V \to C'$ that respects the $M$-criterion and with total misrepresentation zero. First, for every $F \in \mathcal{F}$, set $w(v_F) := c_i$ for an arbitrary chosen element $u_i \in F \cap U'$. Clearly, $r(v_F, c_i) = 0$. So far, the $n$ voters from $V_{\mathcal{F}}$ are assigned to the candidates from $C'$ and it remains to assign the $n(k-1)$ voters from $D$. Since each candidate in $C'$ can represent each dummy voter in $D$ with misrepresentation zero, we can easily extend this assignment so that each candidate from $C'$ is assigned to exactly $n$ voters.

"$\Leftarrow$": Let $C' \subseteq C$ denote a size-$k$ winner set and let $w$ be a mapping from $V$ to $C'$ such that $\sum_{v \in V} r(v, w(v)) = 0$. Since a voter $v_F \in V_{\mathcal{F}}$ can only be represented with cost zero by a candidate $c_i$ if $u_i \in F$, the set $U' := \{u_i \mid c_i \in C'\}$ is a size-$k$ hitting set for $\mathcal{F}$.

This completes the proof for M-MULTIWINNER. It is straightforward to verify that the same construction yields a parameterized reduction for CC-MULTIWINNER. Finally, the W[2]-hardness for the minimax versions follows directly by Observation 1 since the reduction works for $R = 0$. Moreover, NP-hardness directly follows since the reduction can clearly be carried out in polynomial time and containment in NP is obvious. ☐

## 4.2 The Borda Misrepresentation Function

We refine the reduction from the previous subsection to show that also for the Borda misrepresentation function the considered problems are W[2]-hard with respect to the number $k$ of winners as the parameter. However, in contrast to the case of the approval misrepresentation function the reduction does not hold for the case that $R = 0$. Hence, we investigate the parameter total misrepresentation $R$ as well as the composite parameter $(R, k)$ subsequently.

### 4.2.1 The Number of Winners as Parameter

For the Borda misrepresentation function, we also provide a many-one reduction from Hitting Set for M-Multiwinner and then argue that the presented reduction works for the other three problems as well. For CC-Multiwinner W[2]-hardness also directly follows from the NP-hardness reduction (also from Hitting Set) provided by Lu and Boutilier (2011, Thm. 8).[4] Our reduction, however, can deal with the M-criterion and can be adopted to the minimax versions of the two rules; in particular, using some padding of candidates and voters to deal with the M-criterion.

**Theorem 2.** (Minimax) CC-Multiwinner *and* (Minimax) M-Multiwinner *are W[2]-hard with respect to the number $k$ of winners for the Borda misrepresentation function.* Minimax CC-Multiwinner *and* Minimax M-Multiwinner *are NP-complete.*

*Proof.* First, we show W[2]-hardness for M-Multiwinner by a parameterized reduction from Hitting Set. Given an HS-instance $(\mathcal{F} = \{F_1, \ldots, F_n\}, U = \{u_1, \ldots, u_m\}, k)$ build an instance of M-Multiwinner as follows. Let $z := nmk$. The set $C$ of candidates is $C_U \cup B$, where $C_U := \{c_u \mid u \in U\}$ and $B := \{b_i^1, \ldots, b_i^z \mid 1 \le i \le nk\}$. Moreover, the multiset of voters is $V_{\mathcal{F}} \cup D$, where $V_{\mathcal{F}} := \{v_i \mid F_i \in \mathcal{F}\}$ and $D := \{d_1, \ldots, d_{n(k-1)}\}$. For each voter his misrepresentation function is given by his preference list.[5]

Each of the $n$ "set voters" $v_i \in V_{\mathcal{F}}$ has the following preference list:

$$\{c_u \mid u \in F_i\} > b_i^1 > \ldots > b_i^z > \{c_u \mid u \in U \setminus F_i\} > \{b_j^1, \ldots, b_j^z \mid 1 \le j \le nk, j \ne i\}.$$

Finally, for each $i \in \{1, \ldots, n(k-1)\}$, the voter $d_i$ from $D$ has the following preference list:

$$c_1 > c_2 > \ldots > c_m > b_{n+i}^1 > \ldots > b_{n+i}^z > \{b_j^1, \ldots, b_j^z \mid 1 \le j \le nk, j \ne n+i\}.$$

This completes the construction. For the correctness we show the following.

> *Claim.* There is a size-$k$ hitting set for $\mathcal{F}$ if and only if there is a size-$k$ winner set for M-Multiwinner that represents all voters with total misrepresentation at most $z = nmk$.

"$\Rightarrow$": Let $U'$ denote a size-$k$ hitting set for $\mathcal{F}$ and $C' := \{c_u \mid u \in U'\}$. We build a mapping $w : V \to C'$ as follows. First, for every $F_i \in \mathcal{F}$, set $w(v_i) := c_u$ for an arbitrarily chosen element $u \in F_i \cap U'$. Clearly, $r(v_i, c_u) \le m$ since the elements in $F_i$ top the preference list of $v_i$ and $|F_i| \le m$. So far, the $n$ voters from $V_{\mathcal{F}}$ are assigned to the candidates from $C'$.

---

4. The proof is provided in the extended version that appeared at the *Third International Workshop on Computational Social Choice (COMSOC-10)* under the title "Budgeted Social Choice: A Framework for Multiple Recommendations in Consensus Decision Making", see Thm. 6.

5. To improve the readability, we also use sets of candidates in the description of preference lists. Such a set can be fixed in an arbitrary order.

Since each candidate in $C'$ can represent each dummy voter in $D$ with misrepresentation at most $m$, one can extend the mapping so that exactly $n$ voters are assigned to each $c_u \in C'$ with misrepresentation at most $m$ for each voter. Thus, the total misrepresentation of this assignment is at most $nm + nm(k-1) = nmk$.

"$\Leftarrow$": Let $C' \subseteq C$ denote a size-$k$ winner set and $w$ be a mapping from $V$ to $C'$ such that $\sum_{v \in V} r(v, w(v)) \leq mnk$. First, we show that $C'$ can contain no candidate $b_i^j$ from $B$. Every candidate from $B$ can represent at most one voter with misrepresentation value better than $z$. More specifically, if $1 \leq i \leq n$, then $b_i^j$ can represent only the voter $v_i$ with this quality of representation and if, $n < i \leq nk$, then $b_i^j$ can present only the voter $d_i$. Since every candidate from $C'$ must be assigned to exactly $n$ voters and the misrepresentation bound is $z$, we conclude that $C' \cap B = \emptyset$.

It remains to show that $U' := \{u \in U \mid c_u \in C'\}$ is a hitting set for $\mathcal{F}$. A voter $v_i \in V_{\mathcal{F}}$ can be represented by a candidate $c_u \in C_U$ with misrepresentation at most $z$ only if $u \in F_i$ since all candidates $c_{u'}$ with $u' \in U \setminus F_i$ occur in the preference list of $v_i$ after the candidates $b_i^1, \ldots, b_i^z$. Thus, $U'$ is a hitting set for $\mathcal{F}$ of size $k$.

This completes the proof for M-MULTIWINNER. The same construction yields a parameterized reduction for CC-MULTIWINNER based on the same claim. The direction from left to right follows in complete analogy. For the other direction, the only difference is that here a solution set $C'$ of the CC-MULTIWINNER instance might contain a candidate from $B$. However, if there is such a candidate $b_i^j$, then it can represent at most one voter (that is, $v_i$) within the required misrepresentation bound and hence can be replaced by a candidate $c_u \in F_i$ that represents the corresponding voter even better.

For the proof of MINIMAX M-MULTIWINNER and MINIMAX CC-MULTIWINNER, it follows directly from the arguments above that there is a size-$k$ hitting set for $\mathcal{F}$ if and only if there is a set of winners for MINIMAX M/CC-MULTIWINNER consisting of $k$ candidates that represent all voters with maximum misrepresentation at most $R := m - 1$. Hence, W[2]-hardness as well as NP-hardness follow. $\qquad \square$

### 4.2.2 PARAMETER MISREPRESENTATION BOUND

Recall that for the approval misrepresentation function, all four problems are NP-hard even in the fully personalized setting, that is, when $R = 0$. In contrast, for CC-MW and M-MW for the Borda misrepresentation function, we provide polynomial-time algorithms for every constant $R$ while showing that the minimax versions are NP-hard for $R \geq 1$ and polynomial-time solvable for $R = 0$. First, by a simple exhaustive search strategy, one obtains the following.

**Theorem 3.** *For the Borda misrepresentation function,* CC-MULTIWINNER *and* M-MULTIWINNER *are solvable in polynomial time when the misrepresentation bound $R$ is constant.*

*Proof.* In every solution, at most $R$ voters can be represented with misrepresentation greater than 0. Thus, for constant values of $R$, one can try all $O(|V|^R)$ subsets of at most $R$ voters to find a subset $V' \subseteq V$ of voters that are not represented with misrepresentation value zero by an optimal winner set. For each such subset $V'$, for each voter of $v \in V'$, one further tries all possible misrepresentation values from 1 to $R$, that is, one tries $O(R^R)$ possibilities for each $V'$. For each such possibility, the "misrepresentation value" of each voter is determined. Since for Borda there is exactly one candidate that can represent a voter with a specific value, this also implies a corresponding mapping of $V'$ to a set of candidates. Every remaining voter is assigned to the candidate which represents him with misrepresentation value zero. In the case of CC-Multiwinner, it remains to check whether at most $k$ candidates have become representatives and whether the corresponding set of candidates can represent all voters with total misrepresentation at most $R$. In the case of M-Multiwinner one additionally needs to check whether the corresponding assignment satisfies the $M$-criterion. It follows that in both cases an optimal set of $k$ winners can be computed in $O((|V| \cdot R)^R \cdot |V||C|)$ time. $\qquad \square$

Note that Theorem 3 does not imply fixed-parameter tractability with respect to $R$, which remains open in this work. However, we provide fixed-parameter tractability results with respect to the composite parameter $(R, k)$ at the end of this section. Now, we contrast the results for CC-MW and M-MW by showing that the minimax versions become provably more difficult. More specifically, we show the following.

**Theorem 4.** *For the Borda misrepresentation function,* minimax CC-Multiwinner *and* minimax M-Multiwinner *are solvable in polynomial time if the total misrepresentation bound $R = 0$ and are NP-hard for every $R \geq 1$.*

*Proof.* For $R = 0$ polynomial-time solvability follows directly from the fact that every voter $v$ must be assigned to a candidate $c$ with $r(v, c) = 0$ and for the Borda misrepresentation function there is only one such candidate. Hence, one only needs to check if there are less than $k$ such candidates and, for minimax M-Multiwinner whether the corresponding assignment satisfies the M-criterion.

Now, we show NP-hardness for $R = 1$ by a reduction from a special case of Hitting Set. More specifically, Hitting Set is NP-hard even if every set consists of two elements and every element appears in at most three sets (Garey, Johnson, & Stockmeyer, 1974, Thm. 2.4).[6]

Given such a restricted HS-instance $(\mathcal{F} = \{F_1, \ldots, F_n\}, U = \{u_1, \ldots, u_m\}, k)$, build an election as follows. Identify every set from $\mathcal{F}$ with a voter and identify every element from $U$ with a candidate. Moreover, define the following misrepresentation function. For each $F = \{u, v\} \in \mathcal{F}$, let the misrepresentation of voter $F$ be zero for the candidate $u$ and one for the candidate $v$, and the remaining misrepresentation values are assigned arbitrarily to the remaining candidates. Then, the following claim is easy to see.

> *Claim:* There is a hitting set of size $k$ if and only if there is a set of $k$ winners such that the misrepresentation for each voter is at most 1.

This shows the theorem for Minimax CC-MW and $R = 1$. For Minimax M-MW, one can use the following observation showing NP-hardness for an even more restricted setting. It follows directly from the Hitting Set instances constructed in the NP-hardness proof (Garey et al., 1974, Thm. 2.4) that for a yes-instance there is always a hitting set such that every element "hits" either two or three sets from $\mathcal{F}$. More specifically, in case of a yes-instance there is a hitting set $U' \subseteq U$ such that every $u' \in U'$ can be assigned

---

6. The problem is Vertex Cover on cubic graphs.

```
 1  Branch (V', R', C') :
 2  if R' < 0 or |C'| > k  then
 3  │   return "no";
 4  if ∑_{w∈V'}(min_{d∈C'} r(w,d)) ≤ R'  then
 5  │   return "yes";
 6  Consider an arbitrary v ∈ V';
 7  V' := V' \ {v};
 8  for  each c ∈ C with r(v,c) ≤ R'  do
 9  │   R'' := R' - r(v,c) ;
10  │   C'' := C' ∪ {c};
11  │   V'' := V' \ {w ∈ V' | r(w,c) = 0};
12  │   if Branch (V'', R'', C'')  then
13  │   │   return "yes" ;
14  │
15  end
16  return "no";
```

**Algorithm 1:** Branching strategy for CC-MULTIWINNER for Borda misrepresentation functions showing fixed-parameter tractability with respect to the composite parameter $(R, k)$. Initially, the algorithm is invoked with the arguments $(V, R, \emptyset)$. Moreover, $C$ and $k$ are provided as global variables.

either to two or three sets from $\mathcal{F}$ and every set "is hit" by exactly one element. Such a hitting set then one-to-one-corresponds to a winner set fulfilling the M-criterion and hence the theorem also follows for MINIMAX M-MW and $R = 1$. For every $R > 1$, NP-hardness can be proved by similar arguments. Basically, extend the previous construction as follows. For every voter, add $R - 1$ new candidates that are placed at the first $R - 1$ positions for this voter and at a position higher than $R$ for every other voter. Since these new candidates clearly cannot be part of any MINIMAX M-MW solution with misrepresentation bound $R$, one can argue analogously for this case. □

### 4.2.3 COMPOSITE PARAMETER NUMBER OF WINNERS AND MISREPRESENTATION BOUND

In this paragraph, we focus on the scenario that one has a small set of winners that can represent all the voters with small total misrepresentation. This is modeled by the composite parameter $(R, k)$, where $k$ is the number of winners and $R$ is the total misrepresentation. We show that for the Borda misrepresentation function, all four considered problems are fixed-parameter tractable.

**Theorem 5.** *For the Borda misrepresentation function,* CC-MULTIWINNER *and* MINIMAX CC-MULTIWINNER *are fixed-parameter tractable with respect to the composite parameter* $(R, k)$, *where* $k$ *denotes the number of winners and* $R$ *the misrepresentation bound.*

*Proof.* First, we provide a branching strategy for MINIMAX CC-MW. To find a size-$k$ solution we proceed as follows. For an arbitrary voter $v \in V$, branch according to all candidates $c$ with $r(v, c) \leq R$. For each possibility, create a subinstance by deleting each voter $w$ with $r(w, c) \leq R$ from $V$ and recursively solve the corresponding subinstance for $k - 1$. Finally, report whether for at least one subinstance a solution of size $k - 1$ has been found. The recursion stops either if $k < 0$ (reporting "no") or if all voters are represented (reporting "yes").

The correctness of the corresponding algorithm is obvious since a voter $v$ must be represented by a candidate $c$ with $r(v, c) \leq R$. Regarding the running time, one branches into $R + 1$ possibilities for every considered voter and decreases the value of $k$ by one for every subinstance. Hence, the algorithm investigates at most $(R + 1)^k$ possibilities.

We show how to extend this branching strategy to work for CC-MW. The branching recursion is displayed in Algorithm 1 and is invoked with the arguments $(V, R, \emptyset)$. Note that $C$ and $k$ are provided as global variables.

The correctness of Algorithm 1 can be seen as follows. The algorithm first checks whether the misrepresentation bound or the solution size are exceeded (Line 2). Second, the algorithm checks whether the current candidate set is already a winner set, that is, whether all voters are represented and the assignment is within the misrepresentation bound (Line 4). Otherwise, for an arbitrarily chosen voter $v$ (Line 6), the algorithm tries all possible ways of representation without exceeding the misrepresentation bound (Line 8). For each possibility, it decreases $R'$ by the value needed for the representation of $v$ by the corresponding candidate (Line 9). If this possibility implies that a new candidate is added to the current solution, then we can clearly assign all voters that are optimally represented by this candidate to it and hence delete the corresponding voters (Line 11). Finally, we recursively invoke the Branch procedure for the corresponding subinstance (Line 12). Since all possibilities to represent $v$ are considered at least one possibility must lead to a solution (if one exists).

Regarding the running time, we show that before each recursive call (Line 12) the algorithm decreases $R'$ or increases $|C'|$ (or both). In the initial call one has $C' = \emptyset$ and hence $|C'|$ is increased by one (Line 10). For every further call, the only case in which $|C'|$ is not increased is that the considered candidate $c$ is already in the current solution set $C'$. In this case, one cannot have $r(v, c) = 0$ since then $v$ would have been deleted from $V'$ at the point when $c$ has been added to $C'$. Hence, $r(v, c) > 0$ and $R'$ is decreased (Line 9). Since the recursion ends when $R' < 0$ or $|C'| > k$ (Line 2), it follows that the recursion depth is at most $R + k$. Moreover, in each recursive call, one branches according to $R + 1$ possible candidates (Line 8). Hence, the algorithm can be executed in $(R+1)^{R+k} \cdot \mathrm{poly}(n, m)$ time. $\qquad\square$

We remark that the results from Theorem 5 also hold for any instance with misrepresentation functions with nonnegative integer values and $|\{c \in C \mid r(v, c) \leq R\}| \leq R + 1$ for every voter $v \in V$. Moreover, fixed-parameter tractability already follows when $|\{c \in C \mid r(v, c) \leq R\}| \leq f(k, R)$ for any computable function $f$. In contrast, the branching strategy for CC-MW from the Theorem 5 cannot be directly transferred to M-MW since due to the M-criterion one cannot assign a voter to a selected candidate even if it is her best alternative. This means that in an analogous approach the parameter could not be "reduced" and, hence the size of the "search tree" could not be bounded. To show fixed-parameter tractability for M-MW we apply a different approach that employ structural observations based on the M-criterion.

Consider an instance $(C, V, r, R, k)$ of M-MW. Let a *zero-candidate* be a candidate $c \in C$ with $r(v, c) = 0$ for at least one voter $v \in V$.

**Lemma 1.** *In a yes-instance of* M-MULTIWINNER *with the Borda misrepresentation function, there can be at most $R + k$ zero-candidates.*

*Proof.* We apply a proof by contradiction. Assume that there are more than $R + k$ zero-candidates and there is a size-$k$ winner set representing all voters with total misrepresentation at most $R$. For the Borda misrepresentation function, for every voter $v$ there is exactly one candidate $c$ with $r(v, c) = 0$. If $c$ is not part of the winner set, then $v$ contributes by at least one to the total misrepresentation since $r(v, c') \geq 1$ for every $c' \in C \setminus \{c\}$. Since there are more than $R + k$ zero-candidates, more than $R$ of them are not part of a size-$k$ solution. For each of them there is a corresponding voter which is represented by an other candidate of the solution with misrepresentation at least one. Hence, the total misrepresentation is more than $R$; a contradiction. ◻

To make use of the bounded number of zero-candidates, we provide another observation that exploits the M-criterion of a solution.

**Lemma 2.** *Consider an* M-MULTIWINNER *instance with the Borda misrepresentation function. If the number $n$ of voters is greater than $(R + 1)k$, then every size-$k$ set of winners consists of zero-candidates.*

*Proof.* Assume on the contrary that there are more than $(R + 1)k$ voters and a candidate $c$ in the solution set does not represent any of the voters with misrepresentation value zero. Due to the M-criterion and since there are more than $(R + 1)k$ voters, $c$ must represent at least $((R + 1)k)/k = R + 1$ voters with misrepresentation value at least one, respectively. Since the bound for the total misrepresentation is $R$, $c$ cannot be part of a solution. ◻

Based on the two previous lemmas, we show the following.

**Theorem 6.** *For the Borda misrepresentation function, the* M-MULTIWINNER *problem is fixed-parameter tractable with respect to the composite parameter $(R, k)$ where $k$ denotes the number of winners and $R$ the misrepresentation bound.*

*Proof.* An algorithm can be described by distinguishing two cases: $n \leq (R + 1)k$ and $n > (R + 1)k$. If the former, then fixed-parameter tractability follows from Proposition 1 (showing fixed-parameter tractability w.r.t. the number of voters). If the latter, in a yes-instance, by Lemma 1 there are at most $R + k$ zero-candidates and by Lemma 2 the solution has to consist of zero-candidates. After removing all but the zero-candidates, fixed-parameter tractability follows from Proposition 2 (showing fixed-parameter tractability w.r.t. the number of candidates).

Regarding the running time, the first case leads to a running time of $((R + 1)k)^{(R+1)k} \cdot \text{poly}(n, m)$ while the second case can be accomplished in $2^{R+k} \cdot \text{poly}(n, m)$ time. Hence, the theorem follows. ◻

Finally, we show fixed-parameter tractability with respect to $(R, k)$ for Minimax M-MW with the Borda misrepresentation function. The corresponding algorithm is based on the same case distinction as the algorithm for M-MW (Theorem 7). While the basic idea of bounding the number of zero-candidates cannot be transferred from M-MW to Minimax M-MW, the following algorithm for Minimax M-MW works also for M-MW but leads to a worse running time bound for the case that the number of voters exceeds $(R+1)k$. More specifically, for the case $n > (R+1)k$, the exponential running time part is $4^{(R+1)k}$ instead of $2^{R+k}$.

**Theorem 7.** *For the Borda misrepresentation function,* Minimax M-Multiwinner *is fixed-parameter tractable with respect to* $(R, k)$.

*Proof.* Consider a Minimax M-MW instance $(C, V, r, R, k)$ with $r$ being a Borda misrepresentation function. Because the case $R = 0$ is trivial for Minimax M-MW with the Borda misrepresentation function, we assume that $R \geq 1$ in the following.

In the case that $n \leq (R+1)k$, fixed-parameter tractability follows from Proposition 1 (analogously to the proof of Theorem 6). Hence, we consider the case that $n > (R+1)k$. Let $C := \{c_1, \ldots, c_m\}$ and $E_i := \{v \in V \mid r(v, c_i) \leq R\}$ for every $c_i \in C$. Moreover, let $C' := \{c_i \in C : |E_i| \geq \lfloor n/k \rfloor\}$. We show that, first, every solution has to consist of candidates from $C'$ and, second, $|C'| \leq 2(R+1)k$. Then, after removing all but the candidates in $C'$, fixed-parameter tractability follows from Proposition 2.

First, due to the M-criterion at least $\lfloor n/k \rfloor$ voters are assigned to every winning candidate in a solution and hence a candidate $c_i$ with $|E_i| < \lfloor n/k \rfloor$ cannot be part of a winner set.

Second, assume toward a contradiction that $|C'| > 2(R+1)k$. Note that for Borda misrepresentation functions every voter occurs in at most $R+1$ sets from $E_1, \ldots, E_m$. Moreover, since $|E_i| \geq \lfloor n/k \rfloor$ for every $c_i \in C'$

$$n > 2(R+1)k \cdot \lfloor n/k \rfloor \cdot 1/(R+1) > 2k(n/k - 1) = 2n - 2k.$$

Since in the considered case $n > (R+1)k \geq 2k$, this is a contradiction.

Finally, based on Proposition 2, for the case that $n > (R+1)k$, one obtains a running time bound of $4^{(R+1)k} \cdot \text{poly}(n, m)$. $\qquad \Box$

## 5. Single-Peaked Elections

As discussed in the introduction (Subsection 1.4), single-peakedness is a central notion in political science reflecting elections where a single issue dominates preferences of all voters. Let us now formally define this property.

**Definition 5.** Let $V$ be a profile over a set of candidates $C$, and let $\sqsupset$ be a linear order over $C$ (the societal axis). We say that an order $v \in V$ is *compatible* with $\sqsupset$ if for all $c, d, e \in C$ such that either $c \sqsupset d \sqsupset e$ or $e \sqsupset d \sqsupset c$ it holds that

$$\text{pos}_v(c) < \text{pos}_v(d) \implies \text{pos}_v(d) < \text{pos}_v(e). \tag{1}$$

(We remind the reader that positions are counted from the top so that, if $a$ is higher in a linear order than $b$, then the position of $a$ is lower.) We say that $V$ is *single-peaked with*

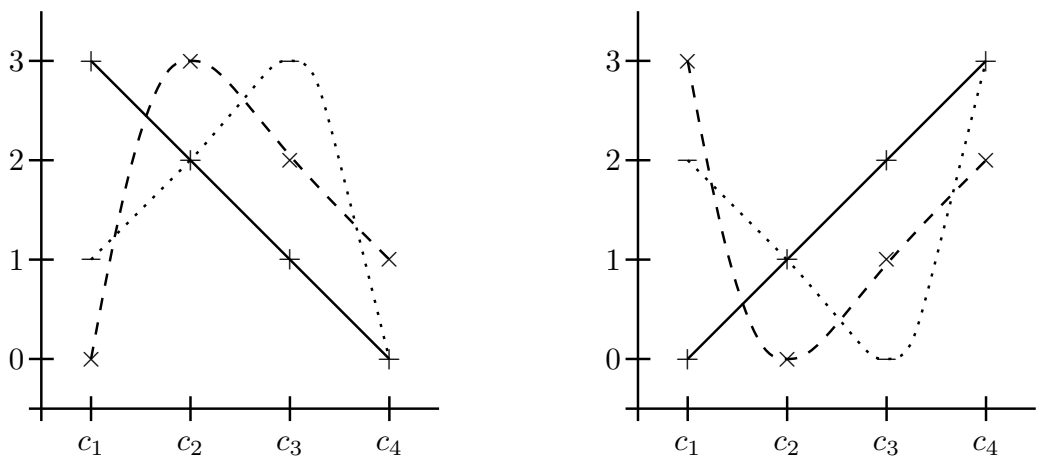

Figure 1: An election consists of three voters with the following preferences: $c_1 > c_2 > c_3 > c_4$, $c_2 > c_3 > c_4 > c_1$, and $c_3 > c_2 > c_1 > c_4$. Its single-peakedness is witnessed by the societal order $c_1 \sqsupset c_2 \sqsupset c_3 \sqsupset c_4$. The diagram on the left-hand side shows, for every voter, the Borda score that each alternative gets from this voter marked by the solid line, dashed line and dotted line, respectively. Note that every "preference order" has one local maximum. If the voters express their Borda misrepresentations values instead, then one obtains the diagram on the right. Here, the misrepresentation function for an arbitrary fixed voter has one local minimum.

*respect to* $\sqsupset$ if for each $i = 1, \ldots, n$ the order $v_i$ is compatible with $\sqsupset$. A profile $V$ is called *single-peaked* if there exists a linear order $\sqsupset$ over $C$ such that $V$ is single-peaked with respect to $\sqsupset$; we will say that $\sqsupset$ *witnesses* the single-peakedness of $V$ and refer to $\sqsupset$ as *societal order*.

**Proposition 3.** *Let $V$ be a single-peaked profile over a set of candidates $C$ witnessed by the societal order $\sqsupset$. Let $r$ be a misrepresentation function for $V$. Then for every triple $\{c_i, c_j, c_k\} \subseteq C$ with $c_i \sqsupset c_j \sqsupset c_k$ or $c_k \sqsupset c_j \sqsupset c_i$ and for every $v \in V$*

$$r(v, c_i) < r(v, c_j) \implies r(v, c_j) \leq r(v, c_k).$$

*Proof.* By the definition of misrepresentation function (see Definition 1) $r(v, c_i) < r(v, c_j)$ implies $\text{pos}_v(c_i) < \text{pos}_v(c_j)$. Now the result follows by (1) and again Definition 1. □

In this section, we investigate the computational complexity of determining proportional representations using Chamberlin and Courant and Monroe methods together with their variants, when the input profile is single-peaked. As discussed in the introduction, when the input profile is single-peaked all voters can be viewed as located on a certain axis where their location is their bliss point. Their most preferred candidate will be either the one closest on the right or the one closest on the left. Without loss of generality we may assume that for each voter her bliss point is the location of one of the candidates (who is then most preferred by her).

It is clear that the misrepresentation function $r(v, c)$ for a single-peaked profile must satisfy the following. If we fix the voter $v$ and change $c$ from one end of the societal axis to the other the value $r(v, c)$ should decrease monotonically to $v$'s most preferred candidate at the bliss point and then increase again monotonically for all candidates beyond the bliss point. That is, for each voter the function expressing the voter's misrepresentation by candidates is single-troughed (that is, has exactly one local minimum) with respect to the order that witnesses single-peakedness of the profile.

Before describing our results, we briefly outline the typical shapes of some prominent misrepresentation functions in single-peaked settings. The Borda misrepresentation function is strictly ascending when moving away from the local minimum in both directions. Moreover, if from the candidate preceding the candidate at the local minimum the misrepresentation function drops by $d > 0$ points, then, for the next $d - 1$ candidates on the other side of the local minimum, the misrepresentation function must ascend in size-one steps.

In contrast, for the approval misrepresentation function, there is exactly one interval of consecutive candidates on the societal axis for whom the misrepresentation is zero while for all remaining candidates outside the interval the misrepresentation is one. For the minimax variants one obtains a similar structure, in the sense that there can be only one interval in which a particular voter can be represented without exceeding the given misrepresentation bound. Note that there is some remote similarity here between the last two cases and preferences over intervals in the aggregating range values model introduced by Farfel and Conitzer (2011).

In the remainder of this section, we provide the following results summarized on Table 2. We show that CC-Multiwinner, Minimax CC-Multiwinner, and Minimax M-Multiwinner for single-peaked elections can be solved in polynomial time for an arbitrary misrepresentation function (Theorem 8, Proposition 4, and Proposition 5, respectively). In contrast to the three aforementioned problems, we present a reduction from an NP-hard version of the Exact 3-Cover problem which shows that M-Multiwinner is NP-hard even when restricted to single-peaked profiles (Theorem 11). However, for the approval misrepresentation function, we still obtain polynomial-time solvability for M-Multiwinner and single-peaked input profiles (Theorem 10). We leave open the computational complexity for M-Multiwinner for the Borda misrepresentation function in the single-peaked case.

## 5.1 (Minimax) CC-Multiwinner

We show that on single-peaked input profiles CC-Multiwinner and Minimax CC-Multiwinner are polynomial-time solvable for an arbitrary misrepresentation function. We first provide a dynamic programming algorithm for the case of CC-Multiwinner. Second, we argue that Minimax CC-Multiwinner can be solved optimally by a greedy algorithm.

### 5.1.1 A Dynamic Programming Procedure for CC-Multiwinner

For CC-Multiwinner the polynomial-time solvability is established by presenting a dynamic programming algorithm leading to the following.

**Theorem 8.** *For a single-peaked input profile and an arbitrary misrepresentation function* CC-Multiwinner *can be solved in* $O(nm^2)$ *time.*

**1 Function SinglePeaked-CC-MW**$(V, C, r, k)$ **Input**: A multiset of
voters $V := \{v_1, \ldots, v_n\}$, a set of candidates $C := \{c_1, \ldots, c_m\}$, a
misrepresentation function $r$, and a positive integer $k$. The voters have
single-peaked preferences according to the societal order $\sqsupset$, where
$c_1 \sqsupset c_2 \sqsupset \ldots \sqsupset c_m$.
**Output**: The minimum total misrepresentation for $k$ winners.

**2 begin**
**3**    **for** $i = 1, \ldots, m$ **do**
**4**        $z(i, 1) := \sum_{v \in V} r(v, c_i)$;
**5**    **end**
**6**    **for** $p = 1, \ldots, m$ **do**
**7**        **for** $i = p + 1, \ldots, m$ **do**
**8**            $d(p, i) := \sum_{v \in V} \max\{0, r(v, c_p) - r(v, c_i)\}$;
**9**        **end**
**10**   **end**
**11**   **for** $i = 2, \ldots, m$ **do**
**12**       **for** $j = 2, \ldots, \min(k, i)$ **do**
**13**          $z(i, j) := \min_{p \in \{j-1, \ldots, i-1\}} (z(p, j-1) - d(p, i))$;
**14**       **end**
**15**   **end**
**16**   **return** $\min_{i \in \{k, \ldots, m\}} (z(i, k))$;
**17 end**

**Algorithm 2:** Dynamic programming algorithm for CC-MULTIWINNER for single-peaked
input profiles.

*Proof.* Throughout the proof we assume that the voters have single-peaked preferences
according to the societal order $\sqsupset$, where $c_1 \sqsupset c_2 \sqsupset \ldots \sqsupset c_m$. For a set $C' \subseteq C$, the
*minimum total misrepresentation* is defined as

$$s(C') = \sum_{v \in V} \min_{c' \in C'} \{r(v, c')\}.$$

We define a dynamic programming table $z$, containing an entry $z(i, j)$ for all $1 \leq i \leq m$
and all $1 \leq j \leq \min(i, k)$. Informally speaking, the entry $z(i, j)$ gives the minimum total
misrepresentation for a set of $j$ winners from $\{c_1, \ldots, c_i\}$ including $c_i$.

The dynamic programming procedure `SinglePeaked-CC-MW` is provided in Algorithm 2.
We show that it solves CC-MULTIWINNER in the claimed running time. Regarding the
correctness, we will show that after the execution of `SinglePeaked-CC-MW` the following
equation is satisfied

$$z(i, j) = \min \left( s(C') \mid C' \subseteq \{c_1, \ldots, c_i\} \wedge |C'| = j \wedge c_i \in C' \right). \quad (2)$$

Then, the minimum total misrepresentation of an optimal size-$k$ winner set is clearly given
by $\min_{i \in \{k, \ldots, m\}} z(i, k)$ (see Line 16).

The proof of Equation 2 follows by induction on $j$. First, we argue that the entries $z(i, 1)$ satisfy Equation (2), yielding the base for induction. To this end, observe that if there is only one candidate $c_i$ in the winner set, then all voters must be assigned to $c_i$, yielding the misrepresentation sum $s(\{c_i\}) = \sum_{v \in V} r(v, c_i)$, see Line 4.

Next, we show that an entry $z(i, j)$ with $j > 1$ (as computed in Line 13) complies with Equation (2) provided that $z(p, j-1)$ does for all $1 \leq p < i$. Consider a set $C^* \subseteq \{c_1, \ldots, c_i\}$ with $c_i \in C^*$ and $|C^*| = j$ such that $s(C^*)$ is minimum among all such sets. We argue that $z(i, j) = s(C^*)$. Let $p < i$ such that $c_p \in C^*$ and $c_\ell \notin C^*$ for all $p < \ell < i$. This implies that $p \geq j - 1$. The crucial observation is as follows. If for a voter $v$ it holds that $r(v, c_p) < r(v, c_p)$, then the single-peakedness implies that $r(v, c_q) \geq r(v, c_p) > r(v, c_i)$ for all $q < p$. Hence, if we consider a set $C''$ of $j - 1$ candidates from $\{c_1, \ldots, c_p\}$ with $c_p \in C''$, then we can assume that the value $r(v, c_p)$ is the contribution of voter $v$ to the total misrepresentation $s(C'')$. Hence, by adding $c_i$ to $C''$ there is an improvement of $r(v, c_p) - r(v, c_i)$ for each voter $v$ with $r(v, c_p) > r(v, c_i)$. For every remaining voter $v$, it holds that $r(v, c_i) \geq r(v, c_p)$ and hence one cannot improve the representation by assigning him to $c_i$. It follows that $s(C^*) = s(C'') - \sum_{v \in V} \max\{0, r(v, c_p) - r(v, c_i)\} = s(C'') - d(p, i)$ and by the induction assumption we have $z(p, j - 1) = s(C'')$. Finally, since the algorithm tries all possible choices of $p$ (see Line 13) we have that $z(i, j) = s(C^*)$.

It is straightforward to verify that the running time of Algorithm 2 is $O(nm^2)$. $\qquad\square$

### 5.1.2 A Greedy Algorithm for Minimax CC-Multiwinner

For single-peaked input profiles, the minimax version of CC-Multiwinner can be solved by a simple greedy algorithm. The basic idea is to iterate over the candidates according to the societal order and to put into the solution the first candidate for whom there is a voter that cannot be represented by the previously selected candidates. The correctness is based on the observation that for each candidate there is a "representation range" (or interval) of consecutive candidates in which the voter must be represented. Thus, we can choose the latest possible candidate that can represent the voter with the representation range that ends first since this candidate is at least as good as every previous candidate.

In other words, the basic combinatorial problem is to cover or stab a given set of intervals (corresponding to the representation ranges of the voters) by $k$ points (corresponding to the candidates). This stabbing problem in turn corresponds to a clique cover problem in interval graphs and can be solved in linear time (Golumbic, 1980). In the next section, we investigate the relationships between the considered voting problems and special (rectangle) stabbing problems in more detail. Here, we conclude with the following.

**Proposition 4.** *For a single-peaked input profile and an arbitrary misrepresentation function* Minimax-CC-Multiwinner *can be solved in $O(nm)$ time.*

### 5.2 (Minimax) M-Multiwinner

We focus on the case when the assignment of the candidates to the winner set satisfies the M-criterion, that is, it is required that each winner represents about the same number of candidates. This additional constraint makes the winner determination more involved. Indeed, we can show that for an integer-valued misrepresentation function M-Multiwinner is NP-hard even if the input profile is single-peaked. On the positive side, we show that M-MULTIWINNER for single-peaked input profiles and the approval misrepresentation function and MINIMAX M-MULTIWINNER for arbitrary misrepresentation functions are polynomial-time solvable. However, the solving strategies (that are also based on dynamic programming) are more intricate than for (MINIMAX) CC-MULTIWINNER. For proving polynomial-time solvability we establish a close relationship to the so-called 1-DIMENSIONAL RECTANGLE STABBING (Even et al., 2008). We start with the polynomial-time algorithms followed by the NP-hardness proof. The computational complexity for M-MULTIWINNER for the Borda misrepresentation function for single-peaked input profiles is left open.

### 5.2.1 M-MULTIWINNER FOR APPROVAL AND MINIMAX M-MULTIWINNER

We use the notation of Even et al. (2008) whenever possible. The input consists of a set $\mathcal{U}$ of horizontal intervals and a set $\mathcal{S}$ of vertical lines with capacity $c(S) \in \{0, \ldots, |\mathcal{U}|\}$ for every line $S \in \mathcal{S}$. Informally, the task is to cover (or stab) all intervals by a minimum number of vertical lines from $\mathcal{S}$, where each line $S$ covers at most $c(S)$ intervals (a vertical line covers a horizontal interval iff they intersect). Since a line $S \in \mathcal{S}$ can cover at most $c(S)$ intervals, one has to specify which interval is assigned to which line in the solution. Let $\mathcal{U}(S)$ denote the set of intervals from $\mathcal{U}$ intersecting with $S \in \mathcal{S}$. An *assignment* is a function $A: \mathcal{S} \to 2^{\mathcal{U}}$, where $A(S) \subseteq \mathcal{U}(S)$. A set $\mathcal{S}' \subseteq \mathcal{S}$ is a *cover* if there is an assignment $A$ with $|A(S)| \leq c(S)$ for all $S \in \mathcal{S}'$ and $\bigcup_{S \in \mathcal{S}'} A(S) = \mathcal{U}$.

> ONE-DIMENSIONAL RECTANGLE STABBING WITH HARD CONSTRAINTS (HARD-1-RS):
> **Input:** A set $\mathcal{U}$ of horizontal intervals and a set $\mathcal{S}$ of vertical lines with capacities $c(S) \in \{0, \ldots, |\mathcal{U}|\}$ for every line $S \in \mathcal{S}$.
> **Task:** Find a minimum-cardinality cover $\mathcal{S}' \subseteq \mathcal{S}$ (and the corresponding assignment).

Now, consider a single-peaked instance of M-MULTIWINNER with $R = 0$ in which every winner represents exactly the same number of voters, that is, $n \bmod k = 0$ for $n$ voters and $k$ winners. In this case, the problem can be "reduced" to HARD-1-RS as follows. For every candidate there is a vertical line according to its position on the societal axis. Since each voter $v$ must be represented by a candidate $c$ with $r(v, c) = 0$ and all the candidates with $r(v, c) = 0$ are ordered consecutively on the societal axis, we can represent each voter by a horizontal interval reaching from the leftmost candidate $c$ with $r(v, c) = 0$ to the rightmost such candidate. Finally, each vertical line is associated with a capacity of $n/k$, which is a whole number. Clearly, there is a solution for the M-MULTIWINNER instance with $R = 0$ if and only if there is a size-$k$ cover for the constructed instance of HARD-1-RS. Note that for the case, when $n \bmod k \neq 0$, this transformation cannot be applied since one does not know whether a candidate line has "capacity" $\lceil n/k \rceil$ or $\lfloor n/k \rfloor$ in an optimal solution.

Even et al. (2008) presented a dynamic programming algorithm for HARD-1-RS with running time $O(|\mathcal{U}|^2 \cdot |\mathcal{S}|^2 \cdot (|\mathcal{U}| + |\mathcal{S}|))$. Since the transformation described above can be easily accomplished in linear time, one directly obtains the following.

**Corollary 1.** *An instance of* M-MULTIWINNER *with a single-peaked profile such that* $n$ mod $k = 0$*, and* $R = 0$ *(and an arbitrary misrepresentation function) can be solved in* $O(n^2 m^2 (n + m))$ *time.*

We show that for single-peaked input profiles, M-MULTIWINNER for the approval misrepresentation function (and arbitrary misrepresentation bound $R$) can be solved in polynomial time. To this end, we show that these instances can be reduced to a version of one-dimensional rectangle stabbing where the goal is to stab a maximum number of horizontal intervals with $k$ vertical lines. More specifically, we introduce the following problem which to the best of our knowledge has not been studied before.

MAXIMUM BALANCED ONE-DIMENSIONAL RECTANGLE STABBING (MAX-BAL-1-RS):

**Input:** A multi-set $\mathcal{U} = \{u_1, \ldots, u_n\}$ of horizontal intervals, a set $\mathcal{S} = \{S_1, \ldots, S_m\}$ of vertical lines, and a positive integer $k$.

**Task:** Find a size-$k$ set $\mathcal{S}' \subseteq \mathcal{S}$ and an assignment $A$ such that each of the following statements hold.

- $|\bigcup_{S \in \mathcal{S}'} A(S)|$ is maximum,
- for every $S \in \mathcal{S}'$, $|A(S)| \leq \lceil n/k \rceil$, and
- $|\{S \in \mathcal{S}' : |A(S)| = \lceil n/k \rceil\}| \leq n$ mod $k$, where $n$ mod $k$ is the remainder on division of $n$ by $k$.

The last two restrictions in the problem description can be considered as saying that we have $k_c = n$ mod $k$ lines with capacity $\lceil n/k \rceil$ and $k_f = k - k_c$ lines with capacity $\lfloor n/k \rfloor$, not specifying which line has which capacity, while for HARD-1-RS there is a specific capacity for every line. The other difference between MAX-BAL-1-RS and HARD-1-RS is that in the latter *all* intervals must be covered by a minimum number of lines whereas in the former the goal is to cover a *maximum* number of intervals with $k$ lines.

We show that the dynamic programming algorithm of Even et al. (2008) for HARD-1-RS can be adapted to work for MAX-BAL-1-RS. To this end, we employ the same decomposition property (stated in Observation 3 below) but the dynamic programming table and the algorithm will be different.

We introduce the following notation to state the dynamic programming. For an interval $u \in \mathcal{U}$, let $l(u)$ denote the left endpoint of $u$ and $r(u)$ denote the right endpoint of $u$ (that is, $l(u) \leq r(u)$). Let $x(S)$ denote the coordinate of line $S \in \mathcal{S}$ and $S(x)$ denote the vertical line associated with a coordinate $x$. For two integers $s$ and $t$ let $[s, t]$ denote the set of all integers $i$ with $s \leq i \leq t$.

For ease of presentation, we assume that the input has the following "normalized" form that can easily be established. First, we assume that all endpoints of the intervals and the coordinates of all lines are integers. Second, we assume that $\{x(S) \mid S \in \mathcal{S}\} = [1, m]$. Third, we assume that the endpoints of all intervals are from $[1, m]$. In what follows, we do not distinguish between a line $S \in \mathcal{S}$ and its coordinate $x(S)$, that is, we identify the lines in $\mathcal{S}$ by the elements of $[1, m]$ (and vice versa). Finally, we assume that the intervals $u_1, u_2, \ldots, u_n$ are ordered so that $l(u_1) \leq l(u_2) \leq \ldots \leq l(u_n)$ (we fix one such ordering).

The algorithm makes use of the fact that there is always an optimal solution that satisfies the "leftmost interval first property" defined as follows (Even et al., 2008).

**Definition 6.** Let $\mathcal{S}' \subseteq \mathcal{S}$ denote a size-$k$ set of lines and let $A$ denote an assignment. We say that $(\mathcal{S}', A)$ has the *leftmost interval first property* if the following holds. Let $S \in \mathcal{S}'$ and let $u_i \in A(S)$. For every $S' \in \mathcal{S}'$ with $l(u_i) \leq S' < S$ and for every $u_j$ with $u_j \in A(S')$, either $j < i$ or $r(u_j) < S$.

Note that the leftmost interval first property is defined with respect to the fixed ordering of the intervals. Any solution can be transformed into an "equivalent" one satisfying the leftmost interval first property by simply swapping the assignments of "conflicting" interval pairs, (see, e.g., Even et al., 2008). Hence, there is always an optimal solution satisfying the leftmost interval first property and one can apply a dynamic programming procedure based on the following decomposition, analogously to the work of Even et al. (2008, Section 2).

**Observation 3.** *Let* $(\mathcal{S}', A)$ *be an optimal solution of* MAX-BAL-1-RS *that satisfies the leftmost interval first property. For any range* $[x_1, x_2] \subseteq [1, m]$, *let* $u_i \in \mathcal{U}$ *be the earliest interval among the intervals covered by a line from* $[x_1, x_2]$ *(that is, for any* $u_j$ *covered by lines in this range we have* $j > i$). *If* $u_i$ *is covered by line* $S \in [x_1, x_2] \setminus \{x_1\}$, *then the right endpoint of all intervals covered by lines in the range* $[x_1, S - 1]$ *are to the left of* $S$.

Basically, Observation 3 is used in the algorithm in the following way. Consider the range $[x_1, m]$ and assume that $x_1$ is the leftmost line of the considered solution. Moreover, assume that $u_i$ is the earliest interval that is covered by a line $S$ from $[x_1, m]$. Then, every interval $u_j$ with $j < i$ will not be covered by the solution, every interval $u_\ell$ with $\ell > i$ and $r(u_\ell) < S$ can only be covered by lines from $[x_1, S - 1]$, and every interval $u_r$ with $S \leq r(u_r)$ can only be covered by lines from $[S, m]$. This implies a decomposition of the instance into two subinstances. The "left" instance contains the intervals $u_g$ with $g > i$ and $r(u_g) < S$ and the "right" instance contains the intervals $u_d$ with $S \leq r(u_d) \leq m$.

**Theorem 9.** MAXIMUM BALANCED ONE-DIMENSIONAL RECTANGLE STABBING *can be solved in* $O(m^3 n^3 k^3)$ *time.*

*Proof.* We use the following definitions to state the dynamic programming algorithm. Let $k_c := n \bmod k$ and $k_f := k - k_c$. For $u_i \in \mathcal{U}$ and for any two coordinates $x_1 \leq x_2$ such that $r(u_i) \in [x_1, x_2]$, let

$$\mathcal{U}(u_i, x_1, x_2) := \{u_j \in \mathcal{U} \mid j \geq i \wedge r(u_j) \in [x_1, x_2]\}.$$

Note that $u_i \in \mathcal{U}(u_i, x_1, x_2)$ and $\mathcal{U} = \mathcal{U}(u_1, 1, m)$.

The algorithm maintains a dynamic programming table with an entry

$$\Pi(u_i, x_1, x_2, k_c', k_f', b) \in \mathbb{N}$$

defined for every $u_i \in \mathcal{U}$, for any two coordinates $x_1 \leq x_2$ such that $r(u_i) \in [x_1, x_2]$ and $x_1 \in u_i$, for each $0 \leq k_c' \leq k_c$ and for each $0 \leq k_f' \leq k_f$ with $k_f' + k_c' \leq k - 1$ and $|[x_1, x_2]| \geq k_c' + k_f' + 1$, and for each $1 \leq b \leq \lceil n/k \rceil$. Informally, the table entry contains the maximum number of intervals from $\mathcal{U}(u_i, x_1, x_2)$ that can be covered by $k_c' + k_f' + 1$ lines from $[x_1, x_2]$ under the assumption that $x_1$ is contained in the solution and covers at most $b$ intervals, $u_i$ is covered by a line from $[x_1, x_2]$, and at most $k_c'$ solution lines different from $x_1$ are assigned to $\lceil n/k \rceil$ intervals. (Formally, the $k_c' + k_f' + 1$ solution lines must satisfy the conditions (C1) to (C6).)

Next, we define subsets of intervals needed for the decomposition into left and right subinstances. For $u_i \in \mathcal{U}$, two coordinates $1 \leq x_1 \leq x_2 \leq m$ such that $r(u_i) \in [x_1, x_2]$, and $x \in [x_1, x_2]$ let

$$\mathcal{U}_l(u_i, x, x_1, x_2) := \begin{cases} \emptyset, & \text{if } x = x_1 \\ \{u_j \in \mathcal{U} \mid j > i \wedge r(u_j) \in [x_1, x-1]\}, & \text{otherwise} \end{cases}$$

and

$$\mathcal{U}_r(u_i, x, x_1, x_2) := \mathcal{U}(u_i, x_1, x_2) \setminus (\mathcal{U}_l(u_i, x, x_1, x_2) \cup \{u_i\}).$$

*Algorithm.* We state the algorithm which will be explained when discussing the correctness below. Basically, the algorithm works in three phases. In a first phase, the dynamic programming table is initialized as follows. For each $u_i \in \mathcal{U}$, for every two coordinates $x_1 \leq x_2$ such that $r(u_i) \in [x_1, x_2]$, $x_1 \in u_i$, and for every integer $b \in [1, \lceil n/k \rceil]$, let

$$\Pi(u_i, x_1, x_2, 0, 0, b) := \min(b, |\{u \in \mathcal{U}(u_i, x_1, x_2) : x_1 \in u\}|). \tag{3}$$

In a second phase, the table is updated. The update of a table entry $\Pi(u_i, x_1, x_2, k'_c, k'_f, b)$ is provided by Algorithm 3, where the order in which the update is invoked is determined by Algorithm 4. In a third phase, the algorithm outputs the maximum value over all $u_i \in \mathcal{U}$ and all $x_1 \in [1, \ldots, m]$ with $x_1 \in u_i$ and $|[x_1, m]| \geq k$ of

$$\max \begin{cases} \Pi(u_i, x_1, m, k_c - 1, k_f, \lceil n/k \rceil) \\ \Pi(u_i, x_1, m, k_c, k_f - 1, \lfloor n/k \rfloor). \end{cases} \tag{4}$$

*Correctness.* We show that in every stage of the dynamic programming an entry contains the value of a "best" assignment of a partial solution for the subinstance with interval set $\mathcal{U}(u_i, x_1, x_2)$ such that six conditions (C1) to (C6) hold. More specifically, we argue that for every $u_i \in \mathcal{U}$, for any two coordinates $x_1 \leq x_2$ such that $r(u_i) \in [x_1, x_2]$ and $x_1 \in u_i$, for each $0 \leq k'_c \leq k_c$ and for each $0 \leq k'_f \leq k_f$ with $k'_f + k'_c \leq k - 1$ and $|[x_1, x_2]| \geq k'_c + k'_f + 1$, and for each $1 \leq b \leq \lceil n/k \rceil$

$$\Pi(u_i, x_1, x_2, k'_c, k'_f, b) = \max |\bigcup_{S' \in \mathcal{S}'} A(S')|$$

over all sets $\mathcal{S}' \subseteq [x_1, x_2]$ and assignments $A : \mathcal{S}' \to 2^{\mathcal{U}(u_i, x_1, x_2)}$ with

- there is an $S \in \mathcal{S}'$ with $u_i \in A(S)$, \hfill (C1)

- $|\mathcal{S}'| = k'_c + k'_f + 1$ , \hfill (C2)

- $x_1 \in \mathcal{S}'$, \hfill (C3)

- $|A(x_1)| \leq b$, \hfill (C4)

- $\forall_{S' \in \mathcal{S}'} |A(S')| \leq \lceil n/k \rceil$, and \hfill (C5)

- $|\{S' \in \mathcal{S}' \setminus \{x_1\} : |A(S')| = \lceil n/k \rceil\}| \leq k'_c.$ \hfill (C6)

**1 Function** UpdateΠ$(u_i, x_1, x_2, k'_c, k'_f, b)$ **begin**

**2**     $M := 0$;

**3**     **if** $b > 1$ **then**

**4**        **for** *every $u_j \in \mathcal{U}(u_i, x_1, x_2) \setminus \{u_i\}$ with $x_1 \in u_j$* **do**

**5**           $M := \max\{M, \Pi(u_j, x_1, x_2, k'_c, k'_f, b-1)\}$;

**6**        **end**

**7**     **else**

**8**        **for** *every $x' = x_1 + 1$ to $x_2$ with $|[x', x_2]| \geq k'_c + k'_f$* **do**

**9**           **for** *every $u_j \in \mathcal{U}(u_i, x_1, x_2) \setminus \{u_i\}$ with $x' \in u_j$* **do**

**10**              **if** $k'_c > 0$ **then**

**11**                 $M := \max\{M, \Pi(u_j, x', x_2, k'_c - 1, k'_f, \lceil n/k \rceil)\}$;

**12**              **if** $k'_f > 0$ **then**

**13**                 $M := \max\{M, \Pi(u_j, x', x_2, k'_c, k'_f - 1, \lfloor n/k \rfloor)\}$;

**14**

**15**           **end**

**16**        **end**

**17**     **for** *every $x = x_1 + 1$ to $r(u_i)$ with $x \in u_i$* **do**

**18**        **for** *all $k^l_c \geq 0$ and $k^r_c \geq 0$ with $k^l_c + k^r_c = k'_c$* **do**

**19**           **for** *all $k^l_f \geq 0$ and $k^r_f \geq 0$ with $k^l_f + k^r_f = k'_f$* **do**

**20**              **if** $|[x_1, x - 1]| \geq k^l_c + k^l_f + 1$ *and* $|[x, x_2]| \geq k^r_c + k^r_f$ **then**

**21**                 $M_l, M_r := 0$;

**22**                 **for** *every $u_j \in \mathcal{U}_l(u_i, x, x_1, x_2)$ with $x_1 \in u_j$* **do**

**23**                    $M_l := \max\{M_l, \Pi(u_j, x_1, x - 1, k^l_c, k^l_f, b)\}$;

**24**                 **end**

**25**                 **for** *every $u_j \in \mathcal{U}_r(u_i, x, x_1, x_2)$ with $x \in u_j$* **do**

**26**                    **if** $k^r_c > 0$ **then**

**27**                       $M_r := \max\{M_r, \Pi(u_j, x, x_2, k^r_c - 1, k^r_f, \lceil n/k \rceil - 1)\}$;

**28**                    **if** $k^r_f > 0$ **then**

**29**                       $M_r := \max\{M_r, \Pi(u_j, x, x_2, k^r_c, k^r_f - 1, \lfloor n/k \rfloor - 1)\}$;

**30**

**31**                 **end**

**32**                 $M := \max\{M, M_l + M_r\}$;

**33**              **end**

**34**           **end**

**35**        **end**

**36**     **end**

**37**     $\Pi(u_i, x_1, x_2, k'_c, k'_f, b) := M + 1$;

**38 end**

**Algorithm 3:** Update step employed by the dynamic programming algorithm for Max-Bal-1-RS presented in the proof of Theorem 9.

```
 1 Main :
 2 for all [x₁, x₂] ⊆ [1, m] in increasing order of x₂ − x₁  do
 3 │   for k'_c = 0, . . . , k_c  do
 4 │   │   for k'_f = 0, . . . , k_f  do
 5 │   │   │   if |[x₁, x₂]| ≥ k'_c + k'_f + 1 and 1 ≤ k'_c + k'_f ≤ k − 1  then
 6 │   │   │   │   for b = 1, . . . , ⌈n/k⌉  do
 7 │   │   │   │   │   for uᵢ ∈ 𝒰 with r(uᵢ) ∈ [x₁, x₂] and x₁ ∈ uᵢ  do
 8 │   │   │   │   │   │   Π(uᵢ, x₁, x₂, k'_c, k'_f, b) := UpdateΠ(uᵢ, x₁, x₂, k'_c, k'_f, b);
 9 │   │   │   │   │   end
10 │   │   │   │   end
11 │   │   │   end
12 │   │   end
13 │   end
14 end
```

**Algorithm 4:** Main loop after the initialization.

For every entry computed in the initialization step (Equation 3), the algorithm stores the maximum value of a "partial solution" $\mathcal{S}' = \{x_1\}$ (satisfies (C2) and (C3)), that covers $u_i$ (C1) (which is possible since $x_1 \in u_i$ and $b \geq 1$), and that satisfies $|A(x_1)| \leq b$ (C4). Clearly, (C5) and (C6) hold as well.

Regarding the update step (see Algorithm 3), let us assume that the values of all entries that are accessed in the update are correct and well-defined (discussed below). To ensure (C1) for an entry $\Pi(u_i, x_1, x_2, k'_c, k'_f, b)$, interval $u_i$ must be covered by one of the lines of $[x_1, x_2]$. We argue that all such possibilities that hold the conditions $C1$ - $C6$ are considered systematically and the current maximum value is stored in the variable $M$. Before going into further details, we observe that Algorithm 3 adds one to the overall maximum value (Line 37) to take into account that $u_i$ has been "newly" covered. This is correct because there is at least one possibility to cover $u_i$ in the considered interval, namely, $x_1$ can always cover $u_1$ since $x_1 \in u_i$ (Line 7 of Algorithm 4) and $b \geq 1$ (Line 6 of Algorithm 4).

Lines 3 to 16 of Algorithm 3 consider the possibility that $x_1$ is used to cover $u_i$. Here, two possibilities are distinguished. The first investigated possibility (Line 3 to Line 7 of Algorithm 3) is that $u_i$ is covered by $x_1$ and $x_1$ can cover at least one more interval, that is, $b > 1$. In this case, we can compute the optimal value based on the value for the subinstance not containing $u_i$ and in which $x_1$ is in the solution but can be assigned to one interval less. To this end, $b$ is decreased by one (that is, (C4) holds) and all possible intervals in $\mathcal{U}(u_i, x_1, x_2) \setminus \{u_i\}$ (Line 4) are checked to be the leftmost covered interval in a corresponding subsolution. Moreover, since $k'_f$ and $k'_c$ remain the same and we assume that conditions (C2), (C5), and (C6) hold for $\Pi(u_j, x_1, x_2, k'_c, k'_f, b-1)$, they also hold for $\Pi(u_i, x_1, x_2, k'_c, k'_f, b)$.

The second investigated possibility (Lines 7 to 16 of Algorithm 3) is that $u_i$ is covered by $x_1$ and $x_1$ can cover at most one interval, that is, $b = 1$. This case can be traced back to the same subinstance without $u_i$ and $x_1$. To access the corresponding possibilities in the dynamic programming table, Algorithm 3 tries all possible lines to be new solution lines (Line 8) and "leftmost" intervals to be covered in the new subinstance (Line 9). Then, it chooses the maximum of assigning a "capacity" of $\lceil n/k \rceil$ (Line 11) or $\lfloor n/k \rfloor$ (Line 13) to $x'$. Note that since $1 \leq k'_c + k'_f$ (Line 5 of Algorithm 4) at least one of these two cases must be possible (if there is at least one further interval to be covered, that is, $\mathcal{U}(u_i, x_1, x_2) \setminus \{u_i\} \neq \emptyset$). Assume that we have a solution corresponding to a table entry $\Pi(u_j, x', x_2, k'_c - 1, k'_f, \lceil n/k \rceil)$ (Line 11) or $\Pi(u_j, x', x_2, k'_c, k'_f - 1, \lfloor n/k \rfloor)$ (Line 13), respectively, and hence, fulfilling the conditions (C1) to (C6) for the corresponding subinstance. Then, clearly adding $x_1$ to this solution and assigning $u_i$ to $x_1$ gives a solution fulfilling all constraints for $\Pi(u_i, x_1, x_2, k'_c, k'_f, b)$ for $b = 1$.

The following loop of Algorithm 3 (Lines 17 to 37) tries all possibilities to cover $u_i$ by a line $x \neq x_1$. If $x \neq x_1$, according to Observation 3, the instance can be divided into two subinstances. All combinations of sizes of the subsolutions are tested by iterating over $k'_c$, $k'_r$, $k'_f$, and $k'^r_f$ (Lines 18 and 19). In Lines 22 to 23 Algorithm 3 computes an optimal solution for the "left" subinstance and in Lines 25 to 29 it computes a solution for the right subinstance obtained after assigning $u_i$ to $x$. The decomposition into the subinstances defined by the interval sets $\mathcal{U}_r(u_i, x, x_1, x_2)$ and $\mathcal{U}_l(u_i, x, x_1, x_2)$ follows directly from Observation 3. Moreover, since $x_1$ is part of the left subinstance with unchanged "capacity bound" $b$, conditions (C1), (C3) and (C4) hold. It remains to show that (C2), (C5) and (C6) hold, that is, in addition to $x_1$ a considered subsolution consists of $k'_c$ lines that are assigned to at most $\lceil n/k \rceil$ intervals and $k'_f$ lines that are assigned to at most $\lfloor n/k \rfloor$ intervals. In any considered possibility, only $x$ is newly specified as a solution line and according to the decomposition (see Observation 3 and the definitions of $\mathcal{U}_l, \mathcal{U}_r$) it is part of the right subinstance. In Lines 27 and 29, Algorithm 3 chooses the maximum between the possibilities that $x$ can be assigned to at most $\lceil n/k \rceil$ or $\lfloor n/k \rfloor$ intervals, respectively, and adapts the values or $k'^r_c$ and $k'^r_f$ accordingly when accessing the corresponding table entries. Hence, all conditions hold.

Now, consider the output of the overall algorithm, see Equation (4). The first maximum function iterates over all intervals $u_i$ and possible leftmost solution lines and hence will find a pair $u_i$ and $x_1$ as follows. The interval $u_i$ is the leftmost interval that is covered by a solution with leftmost interval first property and $x_1$ is the leftmost line of the considered solution. Then, the second maximum function chooses the maximum of the two cases that $x_1$ is assigned to at most $\lceil n/k \rceil$ or $\lfloor n/k \rfloor$ intervals, respectively. Since (C1) to (C6) hold for the corresponding entries, the algorithm outputs the maximum $|\bigcup_{S \in \mathcal{S}'} A(S)|$ overall $\mathcal{S}' \subseteq [1, m]$ and corresponding assignments $A$ fulfilling the further constraints of the definition of MAX-BAL-1RS.

It remains to show that the algorithm only accesses well-defined entries, that is, the accessed entries have been computed before. This is ensured by iterating over the dynamic programming table as described in Algorithm 4. Regarding the computation of $\Pi(u_j, x_1, x_2, k'_c, k'_f, b - 1)$ in Line 5 of Algorithm 3, all parameters values except $b$ are the same as in the current entry. In Line 6 Algorithm 4 iterates over $b$ in increasing order and, hence, $\Pi(u_j, x_1, x_2, k'_c, k'_f, b - 1)$ has been computed before it is accessed. Moreover, the condition $b > 1$ in Line 3 of Algorithm 3 ensures that $\Pi(u_j, x_1, x_2, k'_c, k'_f, b - 1)$ is well defined. In Lines 11 and 13 of Algorithm 3, the accessed range $[x', x_2]$ is smaller than the range from $x_1$ to $x_2$. Since the algorithm iterates over ranges according to increasing size (Line 2 in Algorithm 4) and the other parameter values have also been considered in former iteration loops, the entry has been computed before. In Line 23 of Algorithm 3 the accessed entry has a "range" from $x_1$ to $x - 1 < x_2$ in Lines 27, and 29 the considered range $[x, x_2]$ is also strictly smaller than the range $[x_1, x_2]$. Again, according to Line 2 in Algorithm 4), the entries have been computed before.

*Running time.* Regarding the running time, the update can be accomplished in $O(mk^2n)$ time (see Algorithm 3) by iterating over at most $m$ coordinates of the lines in $\mathcal{S}$ (Line 17), less than $k^2$ cases in Lines 18 and 19, and less than $n$ intervals in the inner loops in Lines 5, 22, and 25, respectively. The overall loop (Algorithm 4) gives an additional factor of $O(m^2n^2k)$: By an appropriate implementation it can be accomplished by iterating over less than $m^2$ coordinate ranges (Line 2), less than $k^2$ possibilities in Lines 4 and 3, $\lceil n/k \rceil \leq (n+1)/k$ values of $b$ (Line 6), and at most $n$ intervals in Line 7. This yields a running time bound of $O(m^2k(n+1)n) = O(m^2n^2k)$. Hence, the overall running time is bounded by $O(n^3k^3m^3)$. $\qquad\square$

An instance with a single-peaked input profile of M-MULTIWINNER for the approval misrepresentation function reduces to MAX-BAL-1-RS by the same transformation as described in Section 5.2.1. We have a vertical line for every candidate, and a horizontal interval for each voter $v$ reaching from the leftmost candidate with $r(v, c) = 0$ and to the rightmost such candidate. The crucial observation is, that minimizing the total misrepresentation for M-MULTIWINNER is equivalent to maximizing the number of voters that are represented by candidates with misrepresentation zero and, hence, to maximizing the number of covered intervals in the MAX-BAL-1-RS instance. Altogether, we arrive at the main result of this section.

**Theorem 10.** M-MULTIWINNER *for the approval misrepresentation function and single-peaked input profiles can be decided in* $O(n^3m^3k^3)$ *time.*

Recall that an instance of MINIMAX M-MULTIWINNER with $R > 0$ can be reduced to an instance of M-MULTIWINNER with $R' = 0$ and approval misrepresentation function by setting for each voter $v$ and each candidate $c$ the misrepresentation value to 0, if $r(v, c) \leq R$, and to 1 otherwise (see Observations 1 and 2). Altogether, we arrive at the following.

**Proposition 5.** *An instance of* MINIMAX M-MULTIWINNER *with a single-peaked profile (and an arbitrary misrepresentation function) can be solved in* $O(n^3m^3k^3)$ *time.*

### 5.2.2 NP-HARDNESS OF M-MULTIWINNER FOR A SINGLE-PEAKED ELECTION

Contrasting the polynomial-time solvability results for the other three considered problems, we show that there is an integer-valued misrepresentation function such that M-MULTIWINNER is NP-complete even restricted to instances with a single-peaked input profile. More specifically, we show that M-MULTIWINNER is NP-hard for single-peaked input profiles and integer-valued misrepresentation functions such that the maximum misrepresentation value of a voter is bounded from above by a polynomial in the number of candidates. Note that for establishing the NP-hardness we have to allow that a voter can assign the same misrepresentation value to several candidates.

The NP-hardness follows by a reduction from a restricted variant of EXACT 3-COVER.

RESTRICTED EXACT 3-COVER (RX3C)
**Input:** A family $\mathcal{S} := \{S_1, \ldots, S_m\}$ of sets over elements $E := \{e_1, \ldots, e_n\}$ such that every set from $\mathcal{S}$ has size 3 and every element of $E$ occurs in exactly three sets.

**Question:** Is there a subset $\mathcal{S}' \subseteq \mathcal{S}$ such that every element of $E$ occurs in exactly one set of $\mathcal{S}'$ and $\bigcup_{S \in \mathcal{S}'} S = E$?

Such a set $\mathcal{S}'$ is called an *exact 3-cover* of $E$. Since in all yes-instances $n$ is a multiple of 3, in what follows we assume that $n$ is divisible by 3. The NP-hardness of RX3C follows from an NP-hardness reduction for the case that every element occurs in at most three subsets (Garey & Johnson, 1979) and a construction to extend this NP-hardness result to the case that every element occurs in exactly three subsets (Gonzalez, 1985).

**Theorem 11.** M-MULTIWINNER *is NP-hard for single-peaked input profiles and an integer-valued misrepresentation function even if the maximum misrepresentation value of every voter is polynomial in the number of candidates (and every winner represents exactly three voters).*

*Proof.* We use the following notation. Consider an RX3C instance $(\mathcal{S}, E)$. For an element $e \in E$ that occurs in the three subsets $S_i$, $S_j$, and $S_k$ with $i < j < k$, we say that the first occurrence of $e$ is in $S_i$, the second occurrence is in $S_j$, and the third occurrence is in $S_k$.

For an RX3C instance $(\mathcal{S}, E)$, define an M-MW instance as follows. The set of candidates is

$$C := E \cup \{s_j \mid S_j \in \mathcal{S}\}$$

and the multiset of voters is

$$V := \{v_i^x \mid e_i \in E \text{ and } x \in \{1,2,3\}\} \cup \{f_i \mid e_i \in E\}.$$

That is, there is a candidate for each element and each subset and there are four voters for each element. Next, we specify the misrepresentation functions of the voters:

| | |
|---|---|
| for $i \in \{1, \ldots, n\}$ | $r(f_i, e_i) := 0$ |
| for $i \in \{1, \ldots, n\}, c \in C \setminus \{e_i\}$ | $r(f_i, c) := 2n^2 + 1$ |
| for $i \in \{1, \ldots, n\}, x \in \{1,2,3\}, 1 \leq z \leq i$ | $r(v_i^x, e_z) := i + z - 1$ |
| for $i \in \{1, \ldots, n\}, x \in \{1,2,3\}, z > i$ | $r(v_i^x, e_z) := 2n^2 + 1$ |
| for $1 \leq j \leq m, x \in \{1,2,3\}$, if the $x$th occurrence of $e_i$ is in $S_j$ | $r(v_i^x, s_j) := 0$ |
| else | $r(v_i^x, s_j) := 1$ |

Finally, set the misrepresentation bound to $R := 2n^2$ and let the number of winners be $k := n/3 + n$. Before showing the correctness of the reduction, we discuss three crucial properties of the construction.

First, we verify that the profile is single-peaked witnessed by the societal order

$$s_1 \sqsupset \cdots \sqsupset s_m \sqsupset e_1 \sqsupset \cdots \sqsupset e_n.$$

For every voter $f_i$ single-peakedness is obvious since his misrepresentation is 0 for one candidate and $2n^2 + 1$ for every other candidate. For every $v_i^x$, within the candidate set $E$, the misrepresentation function decreases monotonously when we move from $e_n$ to $e_1$ along the societal axis: For $z > i$, this is obvious since the misrepresentation remains constant at the value $2n^2 + 1$ and for $z \leq i$ the misrepresentation value is $i + z - 1$ and hence the function clearly assumes smaller values for decreasing values of $z$. This settles the single-peakedness for the "range" from $e_1$ to $e_n$. To see the overall single-peakedness, first note that for $e_1$ the misrepresentation of every $v_i^x$ is at least 1. Then, since the misrepresentation is 1 for all but one of the candidates from $\{s_1, \ldots, s_m\}$ and 0 for the remaining candidates, the single-peakedness for every $v_i^x$ follows.

Second, since there are $4n$ voters and $k = (4n)/3$, exactly three voters have to be assigned to every winning candidate of a solution.

Third, we show that the four voters that can be best represented by candidate $e_i$ are $f_i, v_i^1, v_i^2, v_i^3$. More specifically, we show the following.

**Observation 4.** *For every $e_i$ and $x \in \{1, 2, 3\}$, $r(v_i^x, e_i) < r(y, e_i)$ for every $y \in V \setminus (\{f_i\} \cup \{v_i^1\} \cup \{v_i^2\} \cup \{v_i^3\})$.*

To see the correctness, observe that for every fixed $a \in \{1, \ldots, n\}$, $r(v_a^x, e_a) = 2a - 1$ and for every $x' \in \{1, 2, 3\}$

- for $1 \leq b < a$, $r(v_b^{x'}, e_a) = 2n^2 + 1 > 2a - 1$ ,

- for $a < b \leq n$, $r(v_b^{x'}, e_a) = a + b - 1 > 2a - 1$, and

- for $b \neq a$, $r(f_b, e_a) = 2n^2 - 1 > 2a - 1$.

Now, we show the following.

> *Claim:* There is an exact 3-cover for $(\mathcal{S}, E)$ if and only if there is a set of $k = 4n/3$ candidates that can represent all voters with total misrepresentation $R = 2n^2$ such that exactly three voters are assigned to one candidate.

"$\Rightarrow$" Given an exact 3-cover $\mathcal{S}' \subseteq \mathcal{S}$, we show that the set $\{s_j \mid S_j \in S'\} \cup E$ of candidates is a winning set as required by the claim. The corresponding mapping is as follows.

- For every $1 \leq i \leq n$, the voter $f_i$ is assigned to the candidate $e_i$.

- For $1 \leq i \leq n$ and $x \in \{1, 2, 3\}$, if $e_i$ occurs for the $x$th time in $S_j \in \mathcal{S}'$, then $v_i^x$ is assigned to $s_j$, else $v_i^x$ is assigned to $e_i$.

Since in the exact 3-cover every element is covered exactly once, it follows that every voter is assigned to exactly one candidate and every winning candidate "represents" three voters. More specifically, for the three voters $v_i^1, v_i^2$, and $v_i^3$ corresponding to the three occurrences of the element $e_i$, one of them is represented by the candidate corresponding to the solution set in which $e_i$ occurs and the two other voters by the candidate $e_i$ (the third candidate represented by $e_i$ is $f_i$). It remains to compute the total misrepresentation of this solution. Due to the definition, every candidate $s_j$ represents all three voters with misrepresentation 0. Moreover, every candidate $e_i$ represents $f_i$ with misrepresentation 0 and two voters from $\{v_i^1, v_i^2, v_i^3\}$ with misrepresentation $r(v_i^x, e_i) = i + i - 1 = 2i - 1$ for $x \in \{1, 2, 3\}$. Hence, the total misrepresentation is

$$\sum_{i=1}^{n} 2(2i - 1) = 2n(n + 1) - 2n = 2n^2. \tag{5}$$

511

"⇐" Consider a size-$k$ set $C' \subseteq C$ of winners that represent all voters with total misrepresentation $R = 2n^2$. Since for every voter $f_i$, the only candidate that can represent $f_i$ with misrepresentation at most $R$ is $e_i$, it follows that $E \subseteq C'$. Recall that due to the M-criterion, every candidate must represent exactly three voters. Thus, every candidate $e_i \in E$ must represent two further voters (besides $f_i$). Clearly, a lower bound for the total misrepresentation is achieved in the case when we assign to every $e_i \in E$ two further voters which can be represented by $e_i$ at least as good as any other voters. Due to Observation 4, two such voters are from $\{v_i^1, v_i^2, v_i^3\}$. Moreover, according to Equation 5 the corresponding lower bound for the total misrepresentation matches the total misrepresentation $R = 2n^2$. Since assigning $e_i$ to any voter other than $\{v_i^1, v_i^2, v_i^3\}$ would lead to a strictly higher misrepresentation (Observation 4), this implies that $e_i$ is assigned to exactly two voters from $\{v_i^1, v_i^2, v_i^3\}$. Finally, for every $1 \leq i \leq n$, there remains one voter $v_i^x$ that must be represented by a candidate from $C' \setminus E$ with misrepresentation zero. Since $|C' \setminus E| = n/3$ and a candidate $s_j$ can only represent a voter $v_i^x$ with misrepresentation 0 if the element $e_i$ occurs in $S_j$, i.e., the sets corresponding to the candidates in $C' \setminus E$ must form an exact 3-cover. □

## 6. Conclusion and Outlook

We start with summarizing the relevance of the results of this work. This will be followed by a discussion of closely related problems and models that might be investigated in future research. We conclude with several questions that directly follow from our results.

### 6.1 Relevance of Results

The computation of a set of candidates that "fully proportionally" represent the society has applications in many relevant settings. The main problem with the suggested approaches in the extant literature is that the corresponding combinatorial problems are NP-hard, that is, they cannot be solved efficiently in general. This raises the question whether these approaches despite the theoretically proven advantages (see, e.g., a detailed discussion of those in Brams, 2008) are useless in practice.

One approach is of course to try to escape high complexity by modifying the concept while keeping it still meaningful. In this regard we tried to change the way the total misrepresentation is calculated taking the minimax (or Rawlsian) approach. This appeared not to help in the general case—all problems remain computationally hard—however, it partially helped for single-peaked elections: while the classical Monroe scheme remains NP-hard, its minimax version can be solved in polynomial time.

In general, there are several ways to deal with NP-hard problems. For example, NP-hardness is based on the worst-case analysis and hence one might be able to develop algorithms that work efficiently for most instances. However, although unlikely, it still might happen that the outcome of an election leads to a hard instance. Then, this would lead to the situation of political impasse with unpredictable consequences.

Another common approach to tackle NP-hard problems is to invoke approximation algorithms. While for some scenarios like in the context of resource allocation with sharable items a nearly optimal solution might be sufficient and approximation algorithms are meaningful (Lu & Boutilier, 2011; Skowron, Faliszewski, & Slinko, 2012); for other scenarios, like political elections, the use of approximation algorithms is hard to imagine. A voting rule is a constitutional matter: whatever it is, it must be adhered to. Under the current legislation any candidate or a party can ask for a recount and, if an approximation to a voting rule was used, they may require using another approximation which they can argue give a better representation. It is not hard to imagine prolonged court proceedings on such matters.

Based on the previous discussion, it seems clearly desirable to identify well-specified settings for which an optimal solution can be computed efficiently. This will extend the applicability of the fully proportional representation to such settings. In this regard, we conducted an investigation in two different directions. The first was the class of settings in which some parameters are small (parameterized complexity analysis). The second approach was to restrict attention to single-peaked domains.

Regarding the parameterized complexity of the four studied problems, most of our results are negative (see Table 1). In particular, for the natural and well-motivated parameter the number of winners, the corresponding problems turned out to be W[2]-complete. If, however, in addition there is a winner set that can represent all voters with a small total misrepresentation, three of the four problems become tractable for the Borda misrepresentation function. Moreover, the fixed-parameter tractability results with respect to the number of voters and the number of candidates, respectively, are useful for such restricted settings.

Regarding single-peaked elections, almost all of our results are positive and come with polynomial-time algorithms (see Table 2). A possible critique of this approach is to claim that single-peakedness is in a way an idealized model which is not robust enough. A smallest honest mistake of a voter in filling her ballot may result in election becoming not single-peaked. Also there may be a secondary issue in the election that is also important for some voters which may lead to the election being "almost" single-peaked but not exactly single-peaked. In this regard it would be interesting to investigate how difficult is to find a single-peaked profile "closest" to the given one. For this one might employ techniques of the so-called distance rationalizability approach (Baigent, 1987; Meskanen & Nurmi, 2008; Elkind, Faliszewski, & Slinko, 2010a; Elkind et al., 2010b). It is thus not surprising that near single-peakedness is now starting to be an active area of research (Faliszewski, Hemaspaandra, & Hemaspaandra, 2011; Erdelyi, Lackner, & Pfandler, 2012). Since our algorithms show polynomial-time solvability for the important basic case of single-peakedness, they might be a basis for developing efficient algorithms for such extended settings.

Summarizing, our work contributes to the important topic of making fully proportional representation ideas practical and complements the analysis of this method by Potthof and Brams (1998), Procaccia et al. (2008) and Lu and Boutilier (2011).

## 6.2 Related Problems and Scenarios

Before concluding the work with several open questions, we, first, describe some relations of the considered problems to facility location problems and, second, describe a reasonable alternative multi-winner model. Both topics might also lead to interesting questions for future research.

### 6.2.1 Relations to Facility Location.

A basic scenario for this problem is that a company needs to choose a set of facility locations to serve a set of customers with as little cost to them as possible. Fellows and Fernau (2011) investigated the parameterized complexity of a variant of this problem that is closely related to CC-Multiwinner. Basically, the facility locations can be considered as the set of candidates, the customers as the multiset of voters and the goal is to find a set of facility locations to serve these customers. The only difference is the cost function: In addition to a term that resembles the misrepresentation for every voter (customer), every facility location comes with a certain cost that is required to install the facility.

Similar to our study, Fellows and Fernau (2011) studied the parameter number $k$ of winners/selected facilities locations and the total cost. For the parameter $k$, W[2]-hardness for CC-Multiwinner follows from the reduction given for the facility location problem. Regarding the parameter "total cost", the results of the two papers are not directly comparable. This is due to the fact that the facility location problem stipulates that there is a minimum cost of 1 for serving a customer even at the "best" facility location (which would be an analogue of the condition $r(v, c) \geq 1$ for the misrepresentation function $r$). In this case, the considered problem is fixed-parameter tractable with respect to the total cost. This might not come as a surprise since here the total cost/misrepresentation is at least the number of voters and fixed-parameter tractability with respect to this parameter holds for all four voting problems that we considered (Proposition 2). In contrast, with the condition $r(v, c) \geq 0$ all considered problems are at least W[2]-hard with respect to the total misrepresentation/cost (see Table 1).

The close connection between facility location and multi-winner problems clearly seems to deserve more attention in future work. We remark that analogues of several problems considered in this work might also make sense in the context of the facility location problem. For example, the Monroe model might apply for sets of facilities such that every facility can serve about the same number of customers. Moreover, the single-peaked scenario translates, for example, to the setting that all potential facility locations are along one main street and each resident ranks the cost of using the facility according to the distance from that facility to the place of his residence.

### 6.2.2 Multiset of Candidates Model.

There may be a compromise solution between the two systems of Chamberlin and Courant and Monroe. We may still divide voters into equal or almost equal groups but we may assign the same representative to more than one group of voters. Say, if there are $n$ voters and $k$ representatives are to be elected we may split voters into groups of sizes $\lfloor n/k \rfloor$ and $\lfloor n/k \rfloor + 1$ but allow the same candidate to represent more than one group. Mathematically this would result in selecting not a set of representatives of cardinality $k$ but a multiset of the same cardinality. The classic Monroe (1995) example which considers subscription of newspapers for the common room is in fact a better fit for the multiset model. Indeed, if demand, say for Financial Times, is strong several copies of this newspaper can be subscribed to. We will still need to use weighted voting in the assembly but in this case all weights will be integers.

To illustrate the difference let us consider six people electing a representative assembly of three. Suppose our candidates must come from the set $A = \{a, b, c, d\}$ and the preferences of voters are as follows:

| 4 | 2 |
|---|---|
| a | c |
| b | b |
| c | a |
| d | d |

A set variant of Monroe scheme will give us the set of representatives $\{a, b, c\}$ while from the multiset point of view it is more natural to have a multiset $\{a^2, c\}$ as the answer which could be interpreted to mean that two votes given to $a$ and one to $c$. The multiset point of view seems more natural here, indeed, $b$ does not seem to represent anybody nicely. So the misrepresentation will be nonzero in the set version and zero in the multiset one.

As far as we know the computational complexity for the computation of a winner in the multiset model is unstudied so far. On a first glance, it seems conceivable that the computational complexity for the multiset model lies between the complexity for CC-Multiwinner and M-Multiwinner. This leads to interesting questions such as whether a set of winners according to the multiset model can be computed in polynomial time when the electorate is single peaked.

## 6.3 Open Questions

Several questions arise from this work.

- For CC- and M-Multiwinner for the Borda misrepresentation function we provided algorithms showing polynomial-time solvability for a constant misrepresentation bound $R$. Are these problems fixed-parameter tractable with respect to $R$?

- Is Minimax M-Multiwinner for the Borda misrepresentation function fixed-parameter tractable with respect to the composite parameter $(R, k)$?

- For M-Multiwinner for single-peaked elections we have shown NP-hardness for integer-valued misrepresentation functions. Is the problem fixed-parameter tractable with respect to the number of winners $k$ or/and with respect to the misrepresentation bound $R$?

- Is M-Multiwinner for the Borda misrepresentation function polynomial-time solvable for single-peaked instances?

- Can the results for single-peaked elections be extended to generalized single-peakedness (e.g., as defined by Nehring & Puppe, 2007) or to "almost" single-peaked profiles (in some sense)? This might be of particular interest if the problem of finding the "closest" single-peaked profile to a given one would turn out to be polynomial-time solvable (for some distance on the set of profiles).

## 7. Acknowledgments

Most of the work was done while Nadja Betzler stayed for a research visit at the University of Auckland. Her visit was funded by a fellowship by the Deutscher Akademischer Austausch Dienst (DAAD). Johannes Uhlmann was partially supported by the DFG, project PAWS NI-931/10.

We are very grateful to our referees whose careful and thoughtful reading has significantly improved this paper. Moreover, we like to thank Steven Brams for his interest in this work and valuable discussions and Rolf Niedermeier for his useful comments and support.